\documentclass[aps,pre,preprint,groupedaddress,showpacs]{revtex4-1}
\usepackage[dvips]{graphicx}
\usepackage{textcomp}
\usepackage{amsmath}
\usepackage{amssymb}

\newcommand{\mc}{\multicolumn}

\begin{document}
\title{
\Large\bf Monte Carlo study of the $O(2)$-invariant $\phi^4$ theory with 
a cubic perturbation in three dimensions
}

\author{Martin Hasenbusch}
\email[]{M.Hasenbusch@thphys.uni-heidelberg.de}
\affiliation{
Institut f\"ur Theoretische Physik, Universit\"at Heidelberg,
Philosophenweg 19, 69120 Heidelberg, Germany}
\date{\today}

\begin{abstract}
We study the $2$-component $\phi^4$ model on the simple cubic lattice
in the presence of a cubic, or equivalently, a $\mathbb{D}_4$ invariant 
perturbation. To this end, we perform
Monte Carlo simulations in conjunction with a finite size scaling
analysis of the data. We follow previous work on the $3$-component
case. We study the RG flow from the decoupled Ising fixed point 
into the $O(2)$-invariant one and towards the fluctuation induced
first order transition. To this end we study the behavior of 
phenomenological couplings. At the $O(2)$-invariant fixed point we 
obtain the estimate $Y_4=-0.1118(10)$ of the RG-exponent of the perturbation.  
Note that the small modulus of $Y_4$ means that the RG flow is slow.
Hence, in order to interpret experiments or Monte Carlo simulations 
of lattice models, which are effectively described by the $\phi^4$ model
with a cubic term, we have to consider the RG flow beyond the neighborhood
of the fixed points.
\end{abstract}

\maketitle

\section{Introduction}
We study the renormalization group (RG) flow of the three-dimensional
$\phi^4$ theory with a cubic anisotropy,  
focussing on $N=2$ components of the field. 
For $N=2$, the cubic symmetry is 
equivalent to the dihedral $\mathbb{D}_4$ symmetry. The dihedral group 
contains reflections and rotations, in our case $\mathbb{Z}_4$. 
The problem, for general $N$,
has been studied starting from the early seventies of the last century by using 
field theoretic methods \cite{Wegner,Aha73}. Recently, the $\epsilon$-expansion
has been extended up to 6-loop \cite{epsilon6}. It is
plausible that a cubic perturbation of the $O(N)$ invariant system exists
in real materials due to the underlying crystal lattice structure.
Furthermore the cubic symmetry plays a role in structural transitions
of perovskites, see \cite{AharonyNeu} and references therein. 
Further interest comes from high energy physics. The authors of 
refs. \cite{Arnoldetal97,ArYa97,ArZh97e,ArZh97,Tetradis98,Kari01}
studied the $\phi^4$ theory with a cubic anisotropy and $N=2$ in 
three dimensions as a relatively simple theory, where weak first
order phase transitions occur. The physical motivation of these papers is 
the electro-weak transition in the early universe. Finally, part of the 
phase diagram of the Ashkin-Teller model \cite{AshkinTeller} is effectively
described by the $\phi^4$ theory with cubic anisotropy, see for 
example Refs. \cite{Kadanoff,ArYa97,Schuler}. For the definition of the 
Ashkin-Teller model see appendix \ref{appendixC}.

Field theoretic calculations show that
the cubic perturbation of an $O(N)$-invariant theory in three 
spatial dimensions is irrelevant for $N<N_c$, where $N_c$
is slightly smaller than three. See Refs. \cite{PeVi02,CB_O3} and 
references therein.
This means that for $N=2$ the perturbation is irrelevant.
The corresponding RG-exponent $Y_4 \approx -0.11$ has a rather small modulus. 
For estimates of $Y_4$ obtained by using different methods see
for example refs. 
\cite{PeVi02,Carmona,O234,Debasish,Shao20,O2corrections,Chle22,Cuomo24}.

For instance, to halve the amplitude of the perturbation, we need a scale 
factor $b$, where $b^{-0.11} = 1/2$, hence $b=545.3...$ . In Monte Carlo 
simulations of lattice models or experiments, we can hardly access such 
a large range in the length scale or corresponding temperature ranges
in the neighborhood of the critical temperature.
Hence, analyzing the data, the effect of a finite distance from the 
$O(2)$-invariant fixed point has to be taken into account.
Furthermore, in the case of a finite breaking of the $O(2)$ symmetry, 
it is not clear a priori whether we are within the basin of attraction of the 
$O(2)$-invariant fixed point. For example the $4$-state clock model
can be written as two decoupled Ising models. Hence it undergoes a second order
phase transition but is not in the basin of attraction of the $O(2)$ 
invariant fixed point. For certain ranges of the parameters the 
transition becomes first order.

Here we consider a two component $\phi^4$ model on the simple cubic lattice.
There is a term with cubic symmetry only in the reduced 
Hamiltonian.
We study the model by using Monte Carlo simulations in connection with 
a finite size scaling (FSS) analysis. 
In Ref. \cite{myCubic2} we have studied the analogous problem for three 
components
of the field. For $N=3$, the situation is different, since the cubic 
perturbation is relevant at the $O(3)$ invariant fixed point. As a result, 
a new fixed point, the cubic fixed point arises.
Furthermore, for $N=2$ there is a symmetry that is not present for $N>2$,
which gives rise to considerable simplification. 

The RG flow is traced by the behavior of phenomenological couplings,
or dimensionless quantities,
in finite size scaling. 
Here we borrow ideas from ref. \cite{running}, where the RG flow of a
dimensionless quantity in an asymptotically free theory is discussed.
We obtain an accurate estimate of the RG exponent $Y_4$. Our
numerical results obtained for phenomenological couplings 
for the whole range of the flow could be
compared with those obtained from other models, as for example the 
Ashkin-Teller model, to demonstrate that the same RG flow governs the 
physics at criticality.

The outline of the paper is the following: In Sec. \ref{Model} we
define the model and the observables that we study. 
In Sec. \ref{phasediag} we discuss the qualitative features of the 
RG flow based on the leading order $\epsilon$-expansion. 
In Sec. \ref{RGflow_basic} we outline our FSS method, while in Sec.
\ref{numerics} we discuss the numerical results. In Sec. \ref{FirstOrder}
we focus on the first order phase transitions.
In Sec. \ref{summary} we summarize and compare our estimate of the 
RG exponent $Y_4$ with ones given in the literature.
In the appendix \ref{algorithm} we discuss the Monte Carlo algorithm that is
used in the simulations, and in the appendix
\ref{appendixB} we study the one-component $\phi^4$ model on the simple 
cubic lattice. A brief discussion of the
Ashkin-Teller model is given in the appendix \ref{appendixC}.

\section{The model and observables}
\label{Model}
Here we study the same reduced Hamiltonian and observables as in
ref. \cite{myCubic}. For completeness let us recall the definitions.
We extend the reduced Hamiltonian of the $\phi^4$ model on a simple cubic
lattice, see for example eq.~(1) of ref. \cite{ourHeisen},
by a term proportional to
\begin{equation}
\label{newterm}
 \sum_{a} Q_{4,a a a a}(\vec{\phi}\,) = \sum_{a} \phi_{x,a}^4
-  \frac{3}{N+2}
\left( \vec{\phi}_x^{\,2} \right)^2 \;,
\end{equation}
with cubic symmetry, breaking $O(N)$ invariance.
Note that $Q_4$ is the traceless symmetric combination of four instances of
the field, see for example eq.~(7) of ref. \cite{O234}. We get
\begin{equation}
\label{Hamiltonian}
 {\cal H}(\{\vec{\phi} \,\})= -\beta \sum_{<xy>} \vec{\phi}_x \cdot  \vec{\phi}_y
+ \sum_x \left [ \vec{\phi}_x^{\,2} + \lambda (\vec{\phi}_x^{\,2} -1)^2
+ \mu \left (\sum_{a} \phi_{x,a}^{\,4}
  -\frac{3}{N+2} \left( \vec{\phi}_x^{\,2} \right)^2 \right) \right]
\;,
\end{equation}
where
$\vec{\phi}_x$ is a vector with $N$ real components.
The subscript $a$ denotes the components of the field and
$\{\vec{\phi} \, \}$ is the collection of the fields at all sites $x$.
We label the sites of the simple cubic lattice by
$x=(x_0,x_1,x_2)$, where $x_i \in \{0,1,\ldots,L_i-1\}$. Furthermore,
$<xy>$ denotes a pair of nearest neighbors on the lattice.
In our study, the linear lattice size $L=L_0=L_1=L_2$ is equal in all
three directions throughout. We employ periodic boundary conditions.
The real numbers $\beta$, $\lambda$ and $\mu$ are the parameters of the
model.  Below, discussing the Monte Carlo algorithm, we shall denote 
\begin{equation}
\label{localP}
 V(\phi_x) = \vec{\phi}_x^{\,2} + \lambda (\vec{\phi}_x^{\,2} -1)^2
           + \mu \left (\sum_{a} \phi_{x,a}^{\,4} 
           -\frac{3}{N+2} \left( \vec{\phi}_x^{\,2} \right)^2 \right)
\end{equation}
as local potential.
The partition function 
\begin{equation}
Z = \int [D \phi] \exp[-{\cal H}(\{\vec{\phi} \,\})]
\end{equation}
is only defined as long as contributions from $|\phi_x| \rightarrow \infty$
are suppressed. This is the case as long as the prefactor 
of $(\hat n \cdot \vec{\phi}_x)^4$  in the reduced Hamiltonian remains 
positive for any unit vector $\hat n$.
Let us write $\vec{\phi}_x=r [\cos(\alpha),\sin(\alpha)]$. Then
$(\vec{\phi}_x^2)^2=r^4$ and
$\phi_{x,0}^4+\phi_{x,1}^4=r^4 [\cos^4(\alpha)+\sin^4(\alpha)]$. It follows
$0.5 r^4 \le \phi_{x,0}^4+\phi_{x,1}^4 \le r^4$.  Hence for $\lambda,\mu>0$,
the partition function is defined for $\lambda > \frac{1}{4} \mu$.
For $N>2$ the physics for $\mu>0$ and $\mu<0$ differs. However 
for $N=2$  the two cases are related by a change of variables:
\begin{eqnarray}
\psi_0 &=& 2^{-1/2} (\phi_0 + \phi_1) \\ 
\psi_1 &=& 2^{-1/2} (\phi_0 - \phi_1)  \;,
\end{eqnarray}
which are rotated by $\pi/4$ with respect to the original variables.
It follows $\phi_0 = 2^{-1/2} (\psi_0 + \psi_1)$ and 
$\phi_1 = 2^{-1/2} (\psi_0 - \psi_1)$.
Plugging this into the last term of the reduced Hamiltonian, 
Eq.~(\ref{newterm}), we get
\begin{equation}
\phi_0^4 + \phi_1^4 - \frac{3}{4} (\phi_0^2 + \phi_1^2)^2 =
- \psi_0^4 - \psi_1^4 + \frac{3}{4} (\psi_0^2 + \psi_1^2)^2 \;.
\end{equation}
The term expressed in $\psi$ has the same form as in $\phi$ up to the overall
sign. In the case of the $O(2)$-invariant terms the form obviously stays 
the same, when expressed in $\psi$ instead of $\phi$. 
For this reason we only simulate $\mu > 0$.

\subsection{Decoupled model}
\label{DI_para}
The model decouples into two independent one component systems for 
a vanishing coefficient $\lambda-\frac{3}{N+2} \mu$ of the $(\phi^2)^2$
term in the Hamiltonian, eq.~(\ref{Hamiltonian}). 
Since the term
$\sum_x \vec{\phi}_x^{\,2}$ has the factor $(1-2 \lambda)$ and
$\sum_x \sum_a \phi_{x,a}^{4}$ the factor $\mu = \frac{N+2}{3} \lambda$
in front, a rescaling of the field $\phi_x$ is needed to match with the
Hamiltonian
\begin{equation}
\label{HamiltonianI}
 {\cal H}(\{\phi\})= -\tilde \beta \sum_{<xy>} \phi_x \phi_y
+ \sum_x \left [ \phi_x^{2} + \tilde \lambda (\phi_x^{2} -1)^2
   \right]
\;,
\end{equation}
considered for example in ref. \cite{myPhi4}, where $\phi_x$ is a real
number.
We arrive at the equations
\begin{equation}
 (1-2 \lambda)  = (1-2 \tilde \lambda) \; c \;\; ,\;\;\;\;
\frac{N+2}{3} \lambda = \tilde \lambda \; c^2
\end{equation}
and hence
\begin{equation}
\frac{6}{N+2} \tilde \lambda \; c^2 +  (1-2 \tilde \lambda) \; c -1 =0
\end{equation}
with the solutions
\begin{equation}
\label{conversion}
c =  \frac{-(1-2 \tilde \lambda) \pm \sqrt{(1-2 \tilde \lambda)^2
 +\frac{24}{N+2} \tilde \lambda}}{\frac{12}{N+2} \tilde \lambda} \;\;,
\end{equation}
where we take the positive solution. In Ref. \cite{myPhi4} we find
$\tilde \lambda^* \approx 1.1$, where $\tilde \lambda^*$ 
denotes the value of $\tilde \lambda$, where leading corrections to scaling 
vanish. In appendix \ref{appendixB} we provide an update of this estimate.

The decoupled fixed point is unstable. The corresponding RG-exponent 
is given by 
\begin{equation}
\label{DIexponent}
y_{DI} = \alpha_I y_{t,I} = 2 y_{t,I} - d = 
d - 2 \Delta_{\epsilon, I} = 0.17474944(58) \;,
\end{equation}
where $\alpha_I$ and $y_{t,I}$ are the specific heat and the thermal
RG-exponent of the three-dimensional
Ising universality class, respectively \cite{Sak74,Carmona}. See the 
discussion in section 11.3 of the review \cite{PeVi02}. 
We have used the most recent estimate 
$\Delta_{\epsilon, I}=1.41262528(29)$ obtained by using the conformal 
bootstrap (CB) \cite{CB_Ising_2024}.

\subsection{The observables}
\label{observables}
Dimensionless quantities or phenomenological couplings play a central
role in finite size scaling.
Similar to the study of $O(N)$-invariant models, we study
the Binder cumulant $U_4$, the ratio of partition functions $Z_a/Z_p$ and
the second moment correlation length over the linear lattice size
$\xi_{2nd}/L$. Let us briefly recall the definitions of the observables
and dimensionless quantities that we measure.

The energy of a given field configuration is defined as
\begin{equation}
\label{energy}
 E=  \sum_{<xy>}  \vec{\phi}_x  \cdot \vec{\phi}_y \;\;.
\end{equation}

The magnetic susceptibility $\chi$ and the second moment correlation length
$\xi_{2nd}$ are defined as
\begin{equation}
\label{suscept}
\chi  \equiv  \frac{1}{V}
\left\langle \Big(\sum_x \vec{\phi}_x \Big)^2 \right\rangle \;\;,
\end{equation}
where $V=L^3$ and
\begin{equation}
\label{xi2nd}
\xi_{2nd}  \equiv  \sqrt{\frac{\chi/F-1}{4 \sin^2 \pi/L}} \;\;,
\end{equation}
where
\begin{equation}
F  \equiv  \frac{1}{V} \left \langle
\Big|\sum_x \exp\left(i \frac{2 \pi x_k}{L} \right)
        \vec{\phi}_x \Big|^2
\right \rangle
\end{equation}
is the Fourier transform of the correlation function at the lowest
non-zero momentum. In our simulations, we have measured $F$ for the three
directions $k=0,1,2$ and have averaged these three results.

The Binder cumulant $U_4$ is given by
\begin{equation}
U_{4} \equiv \frac{\langle (\vec{m}^{2})^2 \rangle}{\langle \vec{m}^2\rangle^2},
\end{equation}
where $\vec{m} = \frac{1}{V} \, \sum_x \vec{\phi}_x$ is the
magnetization of a given field configuration. We also consider the ratio
$R_Z\equiv Z_a/Z_p$ of
the partition function $Z_a$ of a system with anti-periodic boundary
conditions in one of the three directions and periodic  ones in the remaining 
two directions and the partition function
$Z_p$ of a system with periodic boundary conditions in all directions.
This quantity is computed by using the cluster algorithm.
For a discussion see Appendix A 2 of ref. \cite{XY1}.

In order to detect the effect of the cubic anisotropy we
study
\begin{equation}
\label{UCdef}
U_C = \frac{\langle \sum_a Q_{4,aaaa}(\vec{m}) \rangle}
     {\langle \vec{m}^{\,2} \rangle^2}  \; .
\end{equation}

In our analysis we need the observables as a function of $\beta$ in
some neighborhood of the simulation point $\beta_s$. To this end we have
computed the coefficients of the Taylor expansion of the observables
up to the third order.

\subsection{Dimensionless quantities for the decoupled system}
\label{DI_R}
In the case of decoupled one-component systems,
$\lambda -\frac{N+2}{3} \mu=0$, we can express
the dimensionless quantities introduced above in terms of their one-component
counterparts.
For example
\begin{equation}
\label{UCDI}
U_{C,DI} = \frac{N-1}{N (N+2)} (U_{4,I} - 3) \;\;,
\end{equation}
where $U_{4,I}$ is the Binder cumulant of the one-component
system. 
The calculation is straight forward, only exploiting that
$\langle m_a^2 m_b^2 \rangle = \langle m_a^2 \rangle  \langle  m_b^2 \rangle$
for $a \ne b$ for the decoupled case.
Furthermore $(Z_a/Z_p)_{DI} = ((Z_a/Z_p)_{I})^N$, $U_{4,DI} =
\frac{1}{N} U_{4,I} + \frac{N-1}{N}$, and
$(\xi_{2nd}/L)_{DI} = (\xi_{2nd}/L)_{I}$, where the subscripts
$DI$ and $I$ indicate the decoupled and the one-component system,
respectively.

\section{Qualitative picture of the RG flow}
\label{phasediag}
\begin{figure}
\begin{center}
\includegraphics[width=14.5cm]{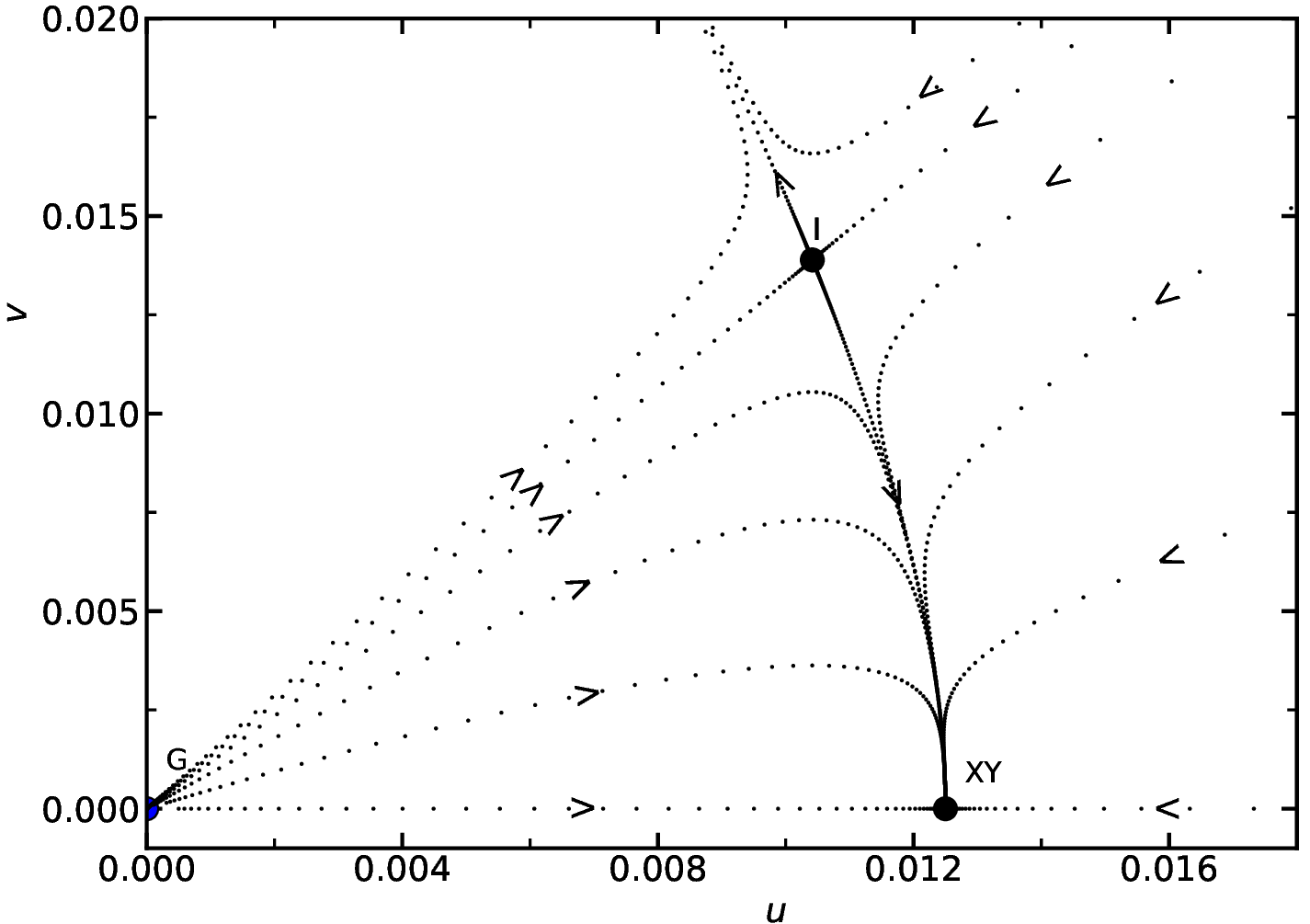}
\caption{\label{rgflow2}
We have numerically integrated the 1-loop flow equations~(\ref{RGeqCardy})
for $N=2$ and
$\epsilon=1$. The fixed points are given as solid circles. The fixed
points are labeled as G (Gaussian), I (decoupled Ising), and 
XY ($O(2)$ invariant).
Selected RG-trajectories are given by dotted lines.
Subsequent dots are separated by a scale factor of $b=2^{1/8}$.
Hence the larger the distance between the dots, the faster the flow.
The arrows indicate the direction of the flow.
}
\end{center}
\end{figure}

In order to get a basic idea of the physics let us discuss the RG flow
obtained from the leading order $\epsilon$-expansion.
The continuum Hamiltonian can be written as
\begin{equation}
\label{ContHamil2mod}
{\cal H} = \int \mbox{d}^d x  \left\{
\frac{1}{2} \sum_{i=1}^{N} [(\partial_{\mu}  \phi_i)^2
+r \phi_i^{2}]  + u \sum_{i,j=1}^{N} \phi_i^2  \phi_j^2 + 
v \left [ \sum_{i=1}^{N} \phi_i^4 - 
\frac{3}{N+2}  \sum_{i,j=1}^{N} \phi_i^2 \phi_j^2 \right] 
\right\}  \;,
\end{equation}
where $\phi_i$ is a real number. Here we write the $\phi^4$ term, following 
ref. \cite{Wegner} in a form similar to eq.~(\ref{Hamiltonian}).
The RG flow, on the critical surface, to leading order in $\epsilon$ becomes
\begin{eqnarray}
\frac{\mbox{d} u}{\mbox{d} l} &=& \epsilon u - 8 (N + 8) u^2 + \frac{ -144 N + 144}{(N+2)^2}  v^2 + ... \;, \nonumber \\
\frac{\mbox{d} v}{\mbox{d} l} &=& \epsilon  v -  96  u  v 
 +  \frac{ 144 - 72 N}{ N + 2}  v^2 + ... \;,
\label{RGeqCardy}
\end{eqnarray}
where $l$ is the logarithm of a length scale.
For $N=2$, $\frac{\mbox{d} v}{\mbox{d} l}$ 
should have only odd powers of $v$ and $\frac{\mbox{d} u}{\mbox{d} l}$
only even powers of $v$. We have generated FIG. \ref{rgflow2}  by 
numerically integrating eqs.~(\ref{RGeqCardy}), setting $\epsilon=1$ and
$N=2$ using selected initial values $(u_0,v_0)$.  The RG flow qualitatively
does not change going to higher orders of the $\epsilon$-expansion 
and apply resummation.
The flow should become even slower. For
$\epsilon=1$ and $N=2$ we get at the $O(2)$ fixed point at $(u,v)=(1/80,0)$
the result
$$
\frac{\mbox{d} v}{\mbox{d} l} = \epsilon  v -  \frac{96}{80}  
 v= -\frac{1}{5} v \;.
$$
Hence $Y_{4,1-loop}=-0.2$ compared with $Y_4 \approx -0.11$. 
We see a hierarchy in the RG flow. Rather rapidly it collapses onto
the line that runs from the decoupled Ising fixed point to the $O(2)$
invariant fixed point. This line further extends from the decoupled Ising 
fixed point towards fluctuation induced first order transitions. In the
following we denote this line as the line of slow RG flow. It determines
the physics at criticality at large scales.

In a first step of our Monte Carlo study, we attempt to identify the
line of the slow flow in the $(\lambda,\mu)$ plane of the parameters 
of the lattice Hamiltonian~(\ref{Hamiltonian}). This step is a 
generalization of finding improved models. In fact, the line ends at
$(\lambda,\mu)=(\lambda^*,0)$, where $\lambda^*$ is the improved value 
of $\lambda$ for the $O(2)$ invariant $\phi^4$ lattice model. Below we
shall also denote the line of the slow flow in the $(\lambda,\mu)$ plane
as improved line.
The slow RG flow itself is then characterized by the behavior of 
$U_C$, eq.~(\ref{UCdef}), at criticality.

\section{Finite size scaling study of the RG flow: theory}
\label{RGflow_basic}
In our study we focus on the one-dimensional flow, running from the 
DI to the $O(2)$-invariant fixed point and from the DI fixed point towards
stronger and stronger first order phase transitions. Here we parameterize the 
flow not in terms of the parameters of the model, but by using the 
dimensionless quantity or phenomenological coupling $U_C$ at criticality.

There are corrections proportional 
to $L^{y_j}$ with $y_j \lessapprox -2$. To deal with these, we simply 
take $L$ sufficiently large, such that they can be ignored, and/or try to 
take them into account at least approximately in the Ans\"atze.  This part
does not appear in the field theoretic discussion.

Beyond that we generalize the idea of improved models. In the case of the 
$O(N)$-invariant $\phi^4$ model, we can tune the parameter $\lambda$
such that leading corrections vanish. Here we are aiming at a line
in the $(\lambda,\mu)$ plane, where one reaches the line of slow flow 
fast, only affected by corrections proportional to $L^{y_j}$ with 
$y_j \lessapprox -2$. Below we shall denote this by the line of slow flow
in the $(\lambda,\mu)$ plane. 

The analysis is split in two parts. In the first step, we locate 
the line of slow flow in the $(\lambda,\mu)$ plane. This is outlined in 
Sec. \ref{slowline} below. Then, by using data generated for values of
$(\lambda,\mu)$ that are close to the line of slow flow, we study the 
flow of $U_C$ at criticality, which is outlined in Sec. \ref{slowflow_theory}. 

\subsection{Locating the line of slow RG-flow in the $(\lambda,\mu)$ plane}
\label{slowline}

In ref. \cite{myCubic2} we have demonstrated in the case of $N=3$  that
$O(N)$-invariant dimensionless quantities $R_i$ can be fitted for a large
range of $(\lambda,\mu)$ by using the Ansatz
\begin{equation}
\label{R3master}
R_i(\beta_c,\lambda,\mu,L) =R_i^*
+ \sum_{m=2}^{m_{max}} c_{i,m} U_C^m(\beta_c,\lambda,\mu,L)
+  r_{i}  w(\lambda,\mu) L^{-\omega}
+ \sum_j a_{i,j}(\lambda,\mu) L^{-\epsilon_j} \;,
\end{equation}
where $R_i^*$ are the fixed point values of the $O(N)$-invariant model.
The motivation of the Ansatz is that there is a hierarchy in the RG-flow.
Very rapidly the flow collapses on a two-dimensional manifold. This 
is taken into account by the last sum in eq.~(\ref{R3master}). In this
sum, $\epsilon_j \gtrapprox 2$.  In the field theoretic setting, 
this is ignored straight away, based on dimensional analysis. 
In the two dimensional manifold, we see again a collapse to a line.
The collapse onto this line is approximated by the term 
$r_{i}  w(\lambda,\mu) L^{-\omega}$. We omit corrections 
proportional to $w(\lambda,\mu)^2 L^{-2 \omega}$, since the values 
of $(\lambda,\mu)$ are chosen such that $w(\lambda,\mu)$ is not that
large. The line of slow RG-flow in the $(\lambda,\mu)$ plane is 
given by the zero of $w(\lambda,\mu)$. In a second step of the 
numerical analysis, where we study the slow RG flow on the one dimensional
manifold, we shall focus on values of $(\lambda,\mu)$ on 
this line.

For $N=2$, the symmetry $\mu \rightarrow -\mu$  implies the corresponding 
symmetry in $U_C$.  Hence we can restrict the sum 
in eq.~(\ref{R3master}) to even values of $m$.   

In our fits we take the numerical estimates 
\begin{eqnarray}
(Z_a/Z_p)^* &=& 0.320380(8) \\
U_4^* &=& 1.242934(10) \\
(\xi_{2nd}/L)^*&=&0.592363(12)
\end{eqnarray}
obtained in Ref. \cite{my3Nclock} as input.

\subsection{Leading corrections}
\label{leading_cor}
In the fits, we deal with $w(\lambda,\mu)$ in two different ways:

\begin{itemize}
\item
First, we take $w$ as a free parameter for each value of $(\lambda,\mu)$
separately. 
\item
Second we use an Ansatz for $w(\lambda,\mu)$ that has only a few free 
parameters.
\end{itemize}
In ref. \cite{myCubic2} we used the parameterizations 
\begin{equation}
 w(\lambda,\mu) = a \;
  [1+c_1 (\lambda-5)] \;
  (\lambda - \lambda^* - b_2 \mu^2 -b_3  \mu^3) \;
\end{equation}
and 
\begin{equation}
 w(\lambda,\mu) = a \;
  [1+c_1 (\lambda-5)] \;
 (\lambda - \lambda^* - b_2 \mu^2 -b_3  \mu^3 -b_4 \mu^4) \;
\end{equation}
of the amplitude of the leading correction for $N=3$. Note that 
$\lambda^* \approx 5$ for $N=3$.  Here we skip odd powers of $\mu$, 
and extend the parameterization:
\begin{equation}
\label{favoritew4}
 w(\lambda,\mu) = a \;
  [1+ c_1 (\lambda-2.15) + c_2 (\lambda-2.15)^2 +g_2 \mu^2 ] \;
(\lambda - \lambda^* - b_2 \mu^2 -b_4  \mu^4)
\end{equation}
and
\begin{equation}
\label{favoritew6}
 w(\lambda,\mu) = a \;
[1+ c_1 (\lambda-2.15) + c_2 (\lambda-2.15)^2 +g_2 \mu^2] \;
(\lambda - \lambda^* - b_2 \mu^2 -b_4  \mu^4 -b_6 \mu^6) \;,
\end{equation}
which results in considerably better fits.  Note that 
in ref. \cite{XY2} $\lambda^*=2.15(5)$ is found for $N=2$. In the fits,
$a$, $c_i$, $g_2$, $b_i$, and $\lambda^*$ are free parameters.
In $(\lambda-2.15)$  we subtract the estimate of $\lambda^*$, with the 
idea that we expand around the improved $O(2)$ invariant fixed point.
Taking a different value, would only change the estimates of $a$, $c_i$ and 
$g_2$, without affecting $\lambda^*$ and $b_i$, which we are interested in.

In the case of the Ising universality class, the most accurate estimate of the
correction exponent is $\omega=0.82968(23)$ obtained by the using the 
CB method \cite{SD16}. In \cite{myClock} we find
$\omega=0.789(4)$ for the XY universality class, which is fully consistent 
with the CB estimate
$\Delta_{s'} = 3.794(8)$  given in table 4 of Ref. \cite{che19}.
In Ref. \cite{DePo20} the authors find $\omega=0.791(8)$ by using the
functional renormalization group (FRG).

\subsubsection{Subleading corrections}
\label{subleading_cor}
In our Ansatz we take into account only subleading corrections with 
$\epsilon_j \approx 2$.  The origin of these corrections are 

\begin{itemize}
\item
The analytic background in the magnetic susceptibility. The corresponding 
correction exponent is $2-\eta$. 
\item
Violations of the spatial rotational symmetry by the simple cubic lattice.
For the Ising and the XY universality class the corresponding leading 
exponent is accurately determined by using CB, $\omega_{NR}=2.022665(28)$, 
Ref. \cite{SD16}, and $\omega_{NR} = 2.02548(41)$, Refs.  
\cite{O2corrections,private}, respectively.
Note that  the value for the XY universality class is only
slightly larger that for the Ising universality class.

\item
The second moment correlation length contains by construction 
a correction with the exponent $2$.  
\end{itemize}

In the fits, we used fixed values of the correction exponents. We 
made no attempt to continuously vary them from the XY to the Ising values.
In the fits discussed below we used the XY values. As check we have replaced
them by the Ising ones. 

In the fits, we assume that the amplitude of the breaking of the 
spatial rotational invariance is constant. This is motivated by the 
fact that this breaking is due to the simple cubic lattice. The 
parameters of the reduced Hamiltonian should have little effect.

The amplitude of the analytic background is parameterized by
\begin{equation}
\label{apm_back}
c_j(\lambda,\mu) = c_{0,j} (1 + b_{j,1} (\lambda-2.15) + g_{j,2} \mu^2)  \;,
\end{equation}
where $j$ is either $\xi_{2nd}/L$ or $U_4$. $Z_a/Z_p$ is not affected 
by the analytic background.

We used two different versions of putting in the subleading corrections
into the fit:
\begin{itemize}
\item
$Z_a/Z_p$:  $d_{Z_a/Z_p} L^{-\omega_{NR}}$ 
\item
$\xi_{2nd}/L$: 
$c_{\xi_{2nd}/L}(\lambda,\mu) L^{\eta-2} + d_{\xi_{2nd}/L} L^{-\omega_{NR}}$ 
\item
$U_4$: $c_{U_4}(\lambda,\mu) L^{\eta-2}$
\end{itemize}
and
\begin{itemize}
\item
$Z_a/Z_p$:  $d_{Z_a/Z_p} L^{-\omega_{NR}}$ 
\item
$\xi_{2nd}/L$: 
$c_{\xi_{2nd}/L}(\lambda,\mu) L^{\eta-2} + d_{\xi_{2nd}/L} L^{-\omega_{NR}}
+ e_{\xi_{2nd}/L}(\lambda,\mu) L^{-2}$ 
\item
$U_4$: $c_{U_4}(\lambda,\mu) L^{\eta-2} + d_{U_4} L^{-\omega_{NR}}$
\end{itemize}
where $e_{\xi_{2nd}/L}(\lambda,\mu)$ is of the form of eq.~(\ref{apm_back}). 

In summary:  The Ans\"atze used in the following are characterized by
\begin{itemize}
\item
The maximal power $m_{max}$ of $U_C$ that is taken into account
\item
The way subleading corrections are parameterized:
Either by using the Ansatz (\ref{favoritew4}) or (\ref{favoritew6}) or
by a free parameter for each value of $(\lambda,\mu)$.
\item
Two different parameterizations for the subleading corrections
\end{itemize}
In the following we shall use the short name $(m_{max},i_w,i_{sub})$ to
characterize the Ansatz. Here $i_w=4,6$ for 
the Ansatz (\ref{favoritew4}) or (\ref{favoritew6}), respectively and 
$i_w=\infty$ for $w$ being a free parameter for each $(\lambda,\mu)$.
And finally $i_{sub}=1,2$ for the two choices to model the subleading 
corrections discussed above.

\subsection{One dimensional flow}
\label{slowflow_theory}
We study the slow RG flow by monitoring the dimensionless quantity $U_C$
at criticality that we denote by $\overline{U}_C$.
The RG flow is characterized by 
\begin{equation}
\label{flowequation}
u(\overline{U}_C) =
\frac{1}{\overline{U}_C}  \frac{\mbox{d} \overline{U}_C}{\mbox{d} \ln L} 
= \frac{\mbox{d} \ln \overline{U}_C}{\mbox{d} \ln L} \;,
\end{equation}
where we have introduced, compared with a $\beta$-function in field theory,
the factor $\frac{1}{\overline{U}_C}$ to lift the zero  at $\overline{U}_C=0$.
In the neighborhood of the $O(N)$-invariant fixed point, where
$\overline{U}_C=0$, the coupling should behave as 
\begin{equation}
\overline{U}_C(L_2)  = \overline{U}_C (L_1) (L_2/L_1)^{Y_4} \;,
\end{equation}
where $Y_4$ is the RG exponent of the cubic perturbation
and  $L_1$ and $L_2$ are two different linear lattice sizes. Here we have 
ignored corrections for simplicity.
Taking the logarithm we get 
\begin{equation}
\ln \overline{U}_C(L_2) - \ln \overline{U}_C(L_1) =Y_4 (  \ln L_2 - \ln L_1 )
\; .
\end{equation}
In the limits $L_2 \rightarrow L_1$ and $\overline{U}_C \rightarrow 0$:
\begin{equation}
\label{relationwithY4}
Y_4 = \frac{\mbox{d} \ln \overline{U}_C}{ \mbox{d} \ln L}  =u(0) \;.
\end{equation}
For a finite argument, we might view $u$ as an effective RG exponent.
In our numerical study, we compute $u(\overline{U}_C)$ by using the 
finite difference approximation
\begin{equation}
\label{finite_diff}
u([\overline{U}_{C,2} + \overline{U}_{C,1}]/2)
= \frac{\ln \overline{U}_{C,2} - \ln \overline{U}_{C,1}}
 {\ln L_2 - \ln L_1} + O(a(u,u',u'',\overline{U}_C) [\ln L_2 - \ln L_1]^2) \;,
\end{equation}
where we have used the abbreviation 
$\overline{U}_{C,i} = \overline{U}_{C}(L_i)$.  The prefactor 
$a(u,u',u'',\overline{U}_C)$ of the correction depends on 
$\overline{U}_C$, $u$ and its derivatives at 
$\overline{U}_C=[\overline{U}_{C,2} + \overline{U}_{C,1}]/2$. In our 
case, where the RG flow is slow, the prefactor $a$ is small.
For $L_2/L_1=2$, which we shall use in the analysis of our data, 
$O(a(u,u',u'',\overline{U}_C) [\ln L_2 - \ln L_1]^2)$ can be ignored 
throughout from the DI to the $O(2)$ invariant fixed point, as we shall
demonstrate explicitly.

In the neighborhood of the DI fixed point we expect
\begin{equation}
\overline{U}_C(L_2)-\overline{U}_{C,DI}^*
= [\overline{U}_{C} (L_1)- \overline{U}_{C,DI}^*]  (L_2/L_1)^{y_{DI}} \;,
\end{equation}
where $\overline{U}_{C,DI}^*$ is the value of $\overline{U}_{C}$ at the 
DI fixed point and $y_{DI}$ is given in Eq.~(\ref{DIexponent}).
Let us define 
\begin{equation}
\label{tildeU}
\tilde u(\tilde U_C) = \frac{1}{\tilde U_C} \frac{\mbox{d} \tilde U_C}
                                                 {\mbox{d} \ln L} \;,
\end{equation}
where $\tilde U_C = \overline{U}_C -\overline{U}_{C,DI}^*$. Analogous to
Eq.~(\ref{relationwithY4}), we get
\begin{equation}
\tilde u(0) = y_{DI} \;.
\end{equation}
Since
\begin{equation}
u(\overline{U}_C) = \frac{\overline{U}_C-\overline{U}_{C,DI}^*}
                         {\overline{U}_C}
  \tilde u(\overline{U}_C-\overline{U}_{C,DI}^*) 
\end{equation}
we have 
\begin{equation}
u(\overline{U}_C) = [y_{DI} + \tilde u'(0) (\overline{U}_C-\overline{U}_{C,DI}^*)
+ ...]
\frac{\overline{U}_C-\overline{U}_{C,DI}^*}
                         {\overline{U}_C}
\end{equation}
in the neighborhood of $\overline{U}_{C,DI}^*$.  In particular 
\begin{equation}
\label{uDI}
u(\overline{U}_{C,DI}^*) = 0 \;\;\; \mbox{and} \;\;  u'(\overline{U}_{C,DI}^*)=
\frac{y_{DI}}{\overline{U}_{C,DI}^*}   \;.
\end{equation}

\section{Finite size scaling study of the RG flow: Numerical results}
\label{numerics}
The Monte Carlo algorithm that is used to generate the data is discussed
in the appendix \ref{algorithm}.

The $O(2)$-invariant $\phi^4$ model, $\mu=0$, had been studied in 
\cite{XY1,XY2}.
In ref. \cite{XY2} $\lambda^*=2.15(5)$ is found.  Close to $\lambda^*$, 
simulations were performed for $\lambda=2.07$, $2.10$, and $2.20$.  The 
estimates of $\beta_c$ are $0.5093835(5)$, 0.5091504(4), and $0.5083355(7)$,
respectively. The data for $\lambda=2.1$ and $2.2$ are used here.

For $\mu>0$, we perform simulations for a number of 
$(\lambda,\mu)$ values, 
where we focus on the line of slow flow in the $(\lambda,\mu)$ plane.
Throughout we simulated the linear lattice sizes 
$L=10$, $11$, ...,$16$, $18$, ..., $24$, $28$, $32$, $40$, ..., $64$.
For $(\lambda,\mu)=(2.13,0.2)$ and $(2.12,0.29)$ we simulated in addition 
$L=72$.

The simulations are performed for $\beta_s \approx \beta_c$, where we 
iteratively improved our estimate of $\beta_c$ with increasing $L$. 
Up to about $L=20$ we performed up to  $6 \times 10^{8}$  
measurements for each $(\lambda,\mu)$ we simulated at. 
For larger $L$, the number of measurements is slowly decreasing with
increasing $L$. For $L=64$ up to $2 \times 10^{8}$ are performed.
In total we used roughly the equivalent of 60 years on a single
core of an Intel(R) Xeon(R) CPU E3-1225 v3 running at 3.20 GHz.

\begin{figure}
\begin{center}
\includegraphics[width=14.5cm]{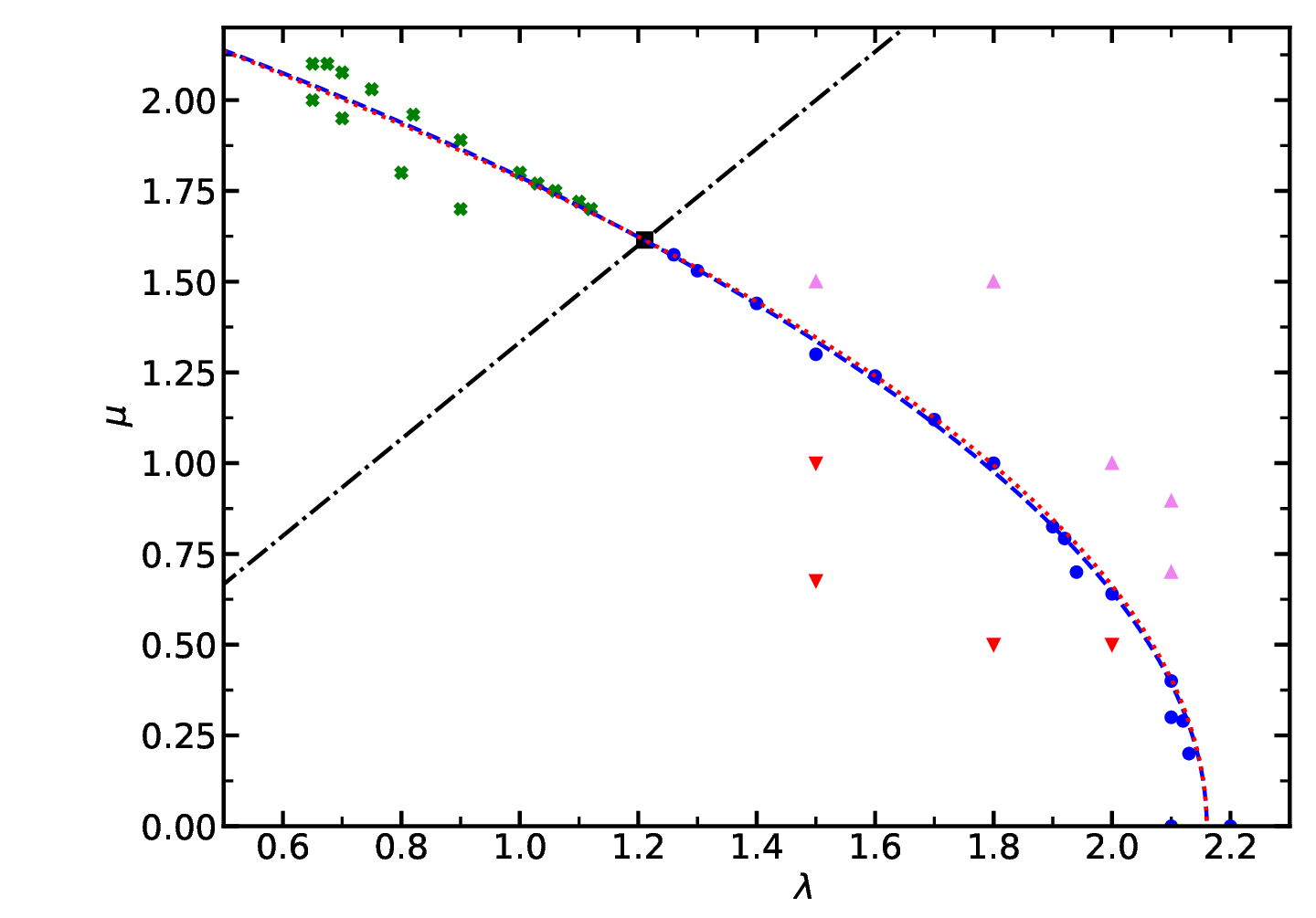}
\caption{\label{improved}
We plot the values of $(\lambda,\mu)$, where we simulated at. 
The decoupled Ising values are given by a dash-dotted black line.
The black square indicates the improved point on this line.
The size of the square gives the error.
The blue dashed line represents our numerical estimate of the 
improved line in the $(\lambda,\mu)$ plane. It is given by
$\lambda=\lambda^* + b_2 \mu^2 +  b_4 \mu^4 +  b_6 \mu^6$, using the 
values given in Eqs.~(\ref{lambdasval},\ref{b2val},\ref{b4val},\ref{b6val}), 
respectively. The red dotted line is chosen such that
it is quadratic in $\mu$ and goes through $\lambda^*$ at $\mu=0$ and the
improved decoupled Ising point. The green 
crosses indicate values, where a first order phase transition occurs.
The blue circles are values with a second order transition close to
the improved line. 
The triangles give values that are further apart from the
improved line. These are used to check the effect of
corrections. For a detailed discussion see the text.
}
\end{center}
\end{figure}

The values of $(\lambda,\mu)$ we simulate at are given in FIG. \ref{improved}.

\subsection{Locating the line of the slow flow: Numerical results}
\label{slowflownum}
Here we build on the discussion of Sec. \ref{slowline}. To organize 
the analysis of the data, 
we have put the values of $(\lambda,\mu)$ into different sets:
\begin{itemize}
\item
set 1: \\  
All values with (expected) second order transitions and ones with 
weak first order transitions:
$(\lambda,\mu)=( 1.12,1.7)$, $(1.1,1.72)$, $(1.06,1.75)$, $(1.03,1.77)$, $(1.0,1.8)$, $(0.9,1.7)$, 
and $(0.9,1.89)$.
\item
set 1m: \\ 
same as set 1, but the values with the largest leading corrections 
$(\lambda,\mu)=(1.5,0.676)$ and $(1.8,1.5)$ and deepest in 
the first order phase, $(0.9,1.7)$ and $(0.9,1.89)$ taken out.
\item
set 2: \\ 
All values with (expected) second order transitions.
\item
set 2m:\\ 
same as set 2, but the two values with the largest leading corrections
taken out.
\item 
set 3: \\ 
Not too close to the decoupled point: $\lambda \ge 1.8$, $\mu \le 1$. 
\end{itemize}

We performed fits by using different versions of the Ansatz~(\ref{R3master}), 
using the different sets of $(\lambda,\mu)$ and varying $L_{min}$, 
the minimal linear lattice size $L$ that is taken into account in the fit.

For each set defined above, we find values of $(m_{max},i_w,i_{sub})$, 
such that the fit is acceptable already for $L_{min}=12$. As one might expect,
for larger sets, larger values of $m_{max}$ are needed.
For example, for set 3, we get acceptable fits for the Ansatz 
$(3,1,1)$ starting from $L_{min}=12$.

\begin{figure}
\begin{center}
\includegraphics[width=14.5cm]{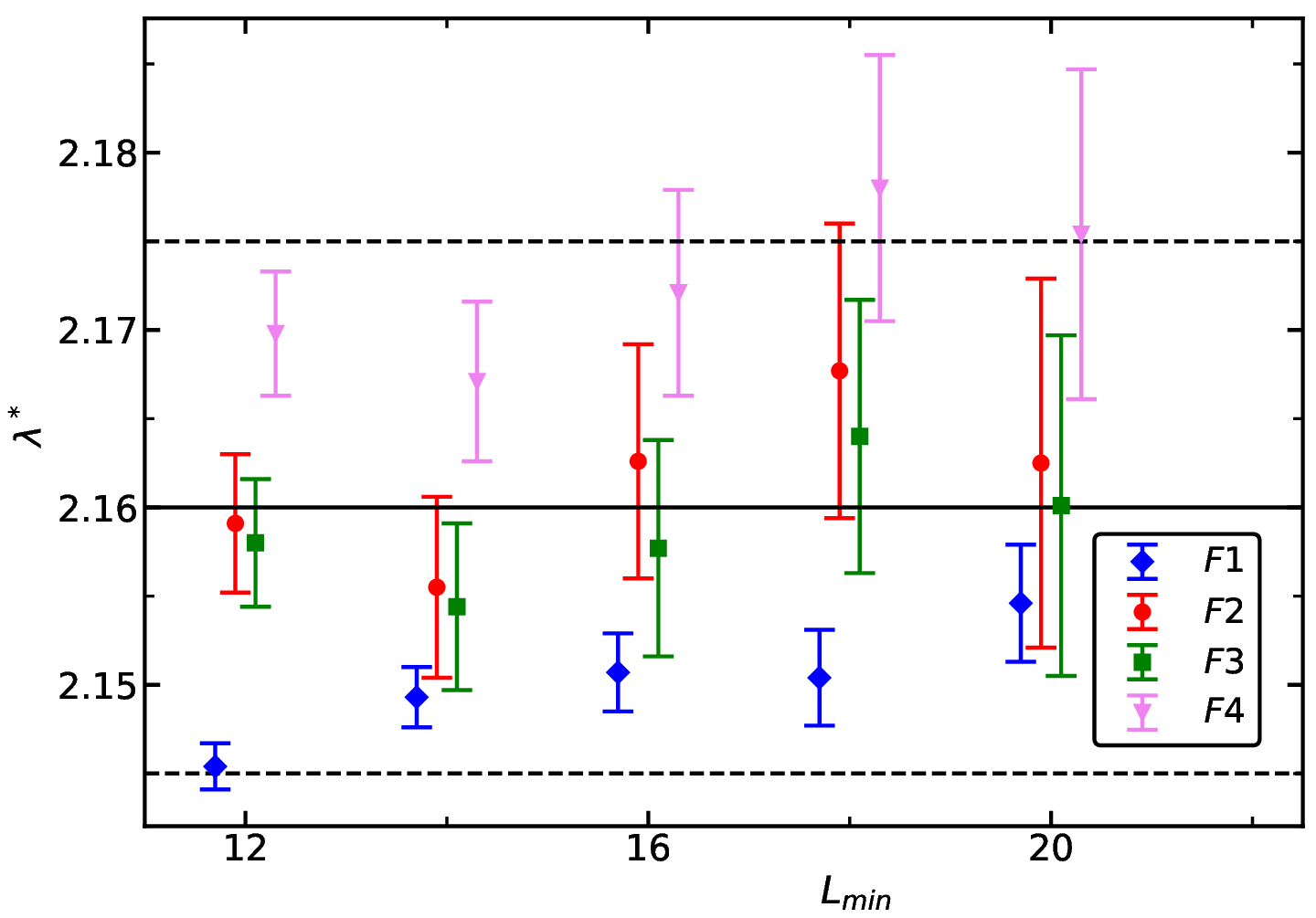}
\caption{\label{lambda_star}
We plot estimates of $\lambda^*$ obtained by using four different Ans\"atze and 
sets of $(\lambda,\mu)$ versus the minimal lattice size $L_{min}$ that is 
taken into account. In particular in the case of fit F1 we take set 3
and the Ansatz is characterized by $(m_{max},i_w,i_{sub})=(6,4,1)$.
For fit F2 we take set 3 and $(m_{max},i_w,i_{sub})=(6,4,2)$.
For fit F3 we take set 2m and $(m_{max},i_w,i_{sub})=(8,4,2)$.
For fit F4 we take set 1m and $(m_{max},i_w,i_{sub})=(10,6,2)$.
The solid line gives our final estimate of 
$\lambda^*$. The dashed lines indicate the preliminary estimate of the 
error. 
The values of $L_{min}$ are slightly shifted to avoid overlap of the symbols.
}
\end{center}
\end{figure}

In Fig. \ref{lambda_star} we plot estimates of $\lambda^*$ obtained 
by using different Ans\"atze and data sets. As our preliminary 
estimate we take $\lambda^*=2.160(15)$. Taking into account the 
uncertainty of the estimates of $R^*_i$ and the correction exponents, 
we arrive at our final estimate
\begin{equation}
\label{lambdasval}
\lambda^*=2.16(3)
\end{equation}

In a similar way, we estimate the central values and errors of 
$b_2$, $b_4$ and $b_6$, eqs.~(\ref{favoritew4},\ref{favoritew6}).
We arrive at
\begin{eqnarray}
 b_2 &=&   -0.39(4) \;, \label{b2val} \\
 b_4 &=&    0.015(35) \;,  \label{b4val} \\
 b_6 &=&   -0.002(6) \;. \label{b6val}
\end{eqnarray}
We notice that the moduli of $b_4$ and $b_6$ are small, compatible with 
zero. 

In FIG. \ref{improved}  we plot $\lambda=\lambda^* + b_2 \mu^2 + b_4 \mu^4 
+b_6 \mu^6$ as approximation of the improved line (line of slow RG flow in 
$(\lambda,\mu)$,)

Next let us discuss the numerical estimates of the coefficients $c_{i,2j}$
that multiply $U_C^{2j}$ in eq.~(\ref{R3master}). 
The coefficients $c_{i,2}$ are quite stable, when varying the maximal
order $j_{max}$ and the data set that is taken into account in the fit.
As our final estimate, compatible with a range of acceptable fits, is
\begin{eqnarray}
c_{Z_a/Z_p,2} &=& - 0.90(2) \\
c_{\xi_{2nd}/L,2}  &=& 1.58(2) \\
c_{U_4,2}  &=& 2.16(3)  \;.
\end{eqnarray}
The amplitude of $c_{i,2j}$ is increasing with $j$. For $j>1$, there 
is a strong dependence on the truncation $j_{max}$  and the set of data 
that is taken into account in the fit.  Hence we abstain from quoting 
final results.

One should note that $j_{max}=1$ is a reasonable approximation up
to the decoupled Ising fixed point. Using the numerical values of 
$R_i^*$ for the $O(2)$ invariant fixed point and decoupled Ising fixed 
point, solving $R_{i,DI} = R_i^* + c_{i,2} U_{C,DI}^2$, we arrive at
$c_{Z_a/Z_p,2} =-0.855$, $c_{\xi_{2nd}/L,2}= 1.666$, and $c_{U_4,2} =1.932$, 
which differ only little from the estimates obtained by the fits with
$j_{max}>1$.  Actually below, to define $\beta_f$
we shall use $c_{Z_a/Z_p,4}=1.4867$, which is chosen such that
$(Z_a/Z_p)^*_{XY}-0.9 U_{C,DI}^{*,2} + c_{Z_a/Z_p,4}  U_{C,DI}^{*,4}
=(Z_a/Z_p)^*_{DI}$.

\subsection{The slow RG flow}
\label{slowflow_numerics}
\begin{figure}
\begin{center}
\includegraphics[width=14.5cm]{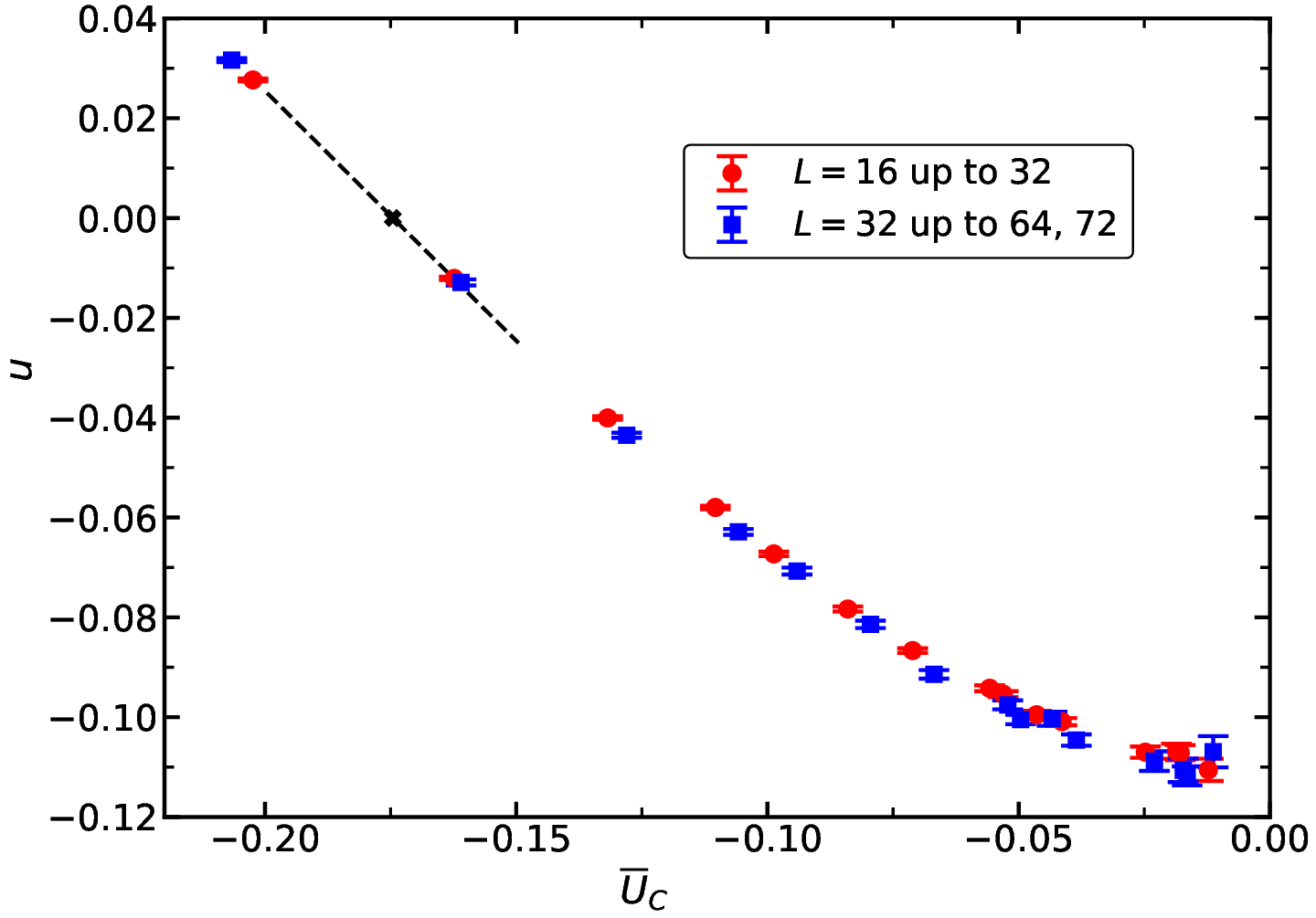}
\caption{\label{flowplot}
We plot numerical estimates of $u(\overline{U}_C)$ obtained for 
two different ranges of lattice sizes $L_{min} \le L \le L_{max}$. 
The black cross gives the decoupled Ising fixed point. The dashed line 
gives the result for the flow in the neighborhood of the 
decoupled Ising fixed point, eq.~(\ref{uDI}).
}
\end{center}
\end{figure}

Here we build on Sec. \ref{slowflow_theory}.
In order to reduce corrections, we focus on $(\lambda,\mu)$ close to 
the improved line in the $(\lambda,\mu)$ plane. We analyze the 
behavior of $\overline{U}_C(\lambda,\mu,L)=U_C(\beta_f(L),\lambda,\mu,L)$, 
where $\beta_f$ is the solution of 
$[Z_a/Z_p](\beta_f,\lambda,\mu,L)-c_{2,Z_a/Z_p} U_C^2(\beta_f,\lambda,\mu,L)
-c_{4,Z_a/Z_p} U_C^4(\beta_f,\lambda,\mu,L) =(Z_a/Z_p)_f$. We take the values
$c_{2,Z_a/Z_p}=-0.9$, $c_{4,Z_a/Z_p} =1.4867$ as discussed above and
$(Z_a/Z_p)_f=0.32038$, which is the estimate of $(Z_a/Z_p)^*$ obtained in
ref. \cite{my3Nclock}. Analogously we could define $\beta_f$ based on 
$\xi_{2nd}/L$. 
Based on eq.~(\ref{finite_diff}), we fit $\overline{U}_C$ by using the Ansatz
\begin{equation}
\overline{U}_C = a L^{u}
\end{equation}
for a certain range of lattice sizes to obtain an estimate of $u$. 
In order to check the effect 
of subleading corrections, we used two ranges of lattice sizes:
Range A: $16 \le L \le 32$  and B: $32 \le L \le 64$. For 
$(\lambda,\mu)=(2.13,0.2)$ and $(2.12,0.29)$ we take $L=72$ instead of
$L=64$ as largest lattice size. As argument of $u$ we take 
$[\overline{U}_C(L_{max}) + \overline{U}_C(L_{min})]/2$.  
In Fig. \ref{flowplot} our numerical results are plotted.

We have fitted the estimates of $u$ by using the Ans\"atze
\begin{equation}
\label{ufit2}
 u=u_0 + u_2 \overline{U}_C^2
\end{equation} 
and 
\begin{equation}
\label{ufit4}
 u=u_0 + u_2 \overline{U}_C^2 + u_4 \overline{U}_C^4 \;.
\end{equation}
We took all available data down to a minimal $\overline{U}_C$ into account.
In the case of the Ansatz~(\ref{ufit2}) we get an acceptable 
fit down to $\overline{U}_C =-0.11036$ for the range $16 \le L \le 32$
with the results $u_0=-0.10761(34)$ and $u_2=4.103(42)$.  For a minimal 
value $\overline{U}_C =-0.0987417$ we get instead 
$u_0=-0.10796(39)$ and $u_2=4.195(67)$. Using the Ansatz~(\ref{ufit4}) 
we get an acceptable  fit even when including all our data 
up to the decoupled Ising point. For example, for a minimal
value $\overline{U}_C =-0.16235475$ we get $u_0=-0.10858(37)$, 
$u_2=4.551(65)$ and $u_4=-33.9(2.2)$.  We conclude $u_0=-0.1082(7)$. 
Following the same procedure for the ranges $32 \le L \le 64$ or 
$32 \le L \le 72$
we get $u_0 = -0.1109(7)$. Assuming
that the difference is mainly caused by subleading correction that decay 
like $L^{-2}$  we arrive at our final estimate $u_0 =-0.1118(10)$.

Finally we checked whether for the finite difference 
approximation~(\ref{finite_diff}) with $L_2/L_1=2$ the error can be ignored.
To this end we numerically integrated $\overline{U}_C$ taking $u$ given 
by Eq.~(\ref{ufit4}) with the numerical values
$u_0=-0.11$, $u_2=4.55$, and $u_4=-34$. Taking these values of $\overline{U}_C$
we compute $\hat u$ by using Eq.~(\ref{finite_diff}) and $L_2/L_1=2$.
We find that $\hat u -u$ has one minimum with the value 
$\approx -2.8 \times 10^{-6}$ at $\overline{U}_C \approx -0.049$ and one maximum
with the value $\approx 1.6 \times 10^{-5}$ at $\overline{U}_C \approx -0.139$. 

\subsection{Effective exponent $\nu$ of the correlation length}
\label{yteff}
As suggested for example in ref. \cite{AharonyNeu}, we determine an effective
RG exponent $y_{t,eff}=1/\nu_{eff}$, where $\nu$ is the 
critical exponent of the correlation length.

To this end, we fitted the slope of dimensionless quantities
\begin{equation}
\overline{S}_R =\left . \frac{\partial R}{\partial \beta} \right |_{\beta=\beta_f} \;,
\end{equation}
where $\beta_f$ is defined as in the section above,
with the Ansatz
\begin{equation}
 \overline{S}_R = a L^{y_{t,eff}} (1 + c L^{-\epsilon}) \;,
\end{equation}
where we have taken $\epsilon = 2.02548$. Note that for example 
$\epsilon =2$ leads to virtually the same results. We take into account
linear lattice sizes $L_{min} \le L \le L_{max}$. Throughout  
$L_{min}=16$, while $L_{max}=64$, except for $(\lambda,\mu)=(2.13,0.2)$
and $(2.12,0.29)$, where $L_{max}=72$.
In most cases, these fits have an acceptable goodness of the fit. 
Let us just view this procedure as definition of an effective 
RG exponent $y_{t,eff}$. 

\begin{figure}
\begin{center}
\includegraphics[width=14.5cm]{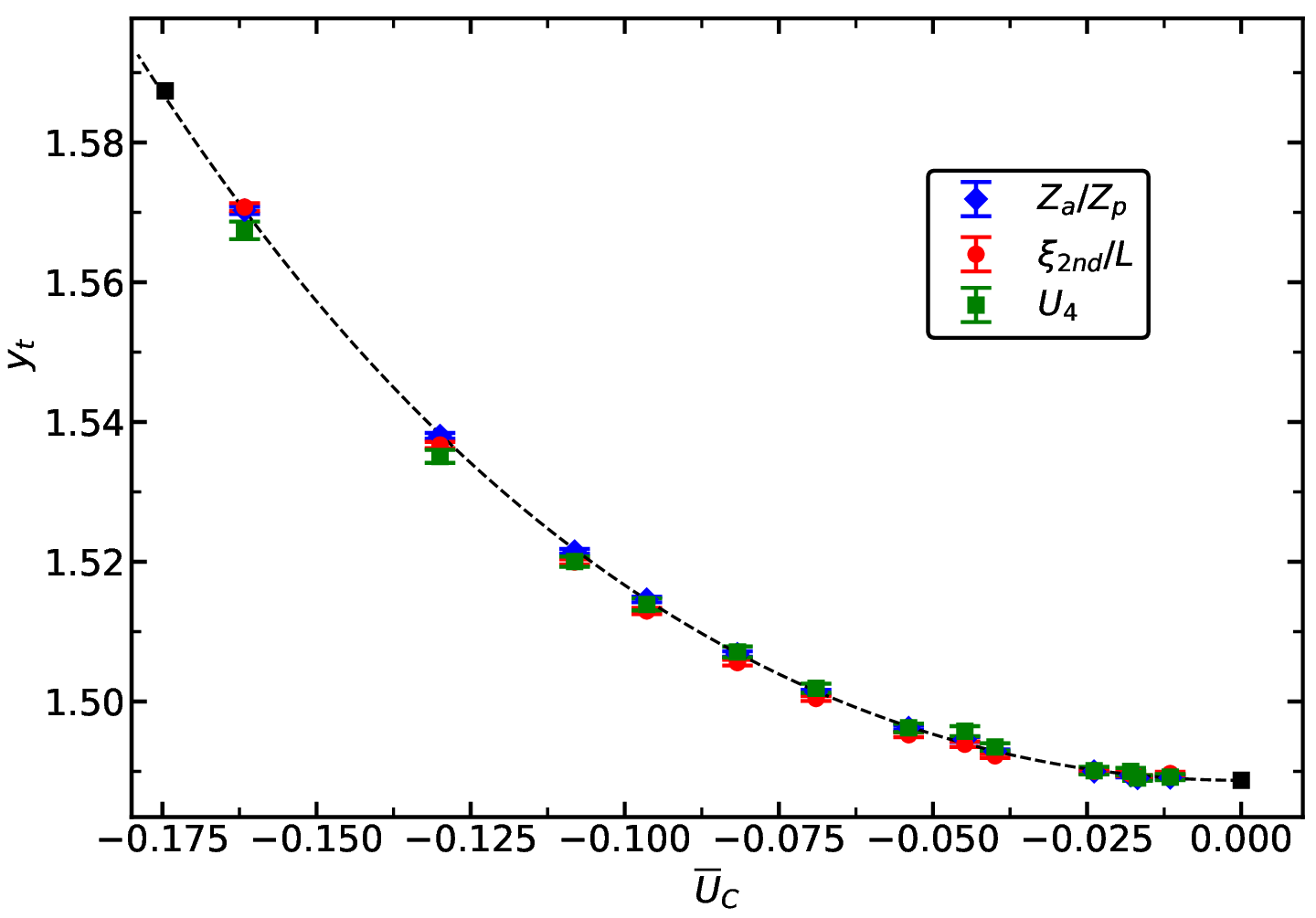}
\caption{\label{yteffplot}
We plot numerical estimates of $y_{t,eff}$  versus $\overline{U}_C$. 
At $\overline{U}_C=0$ we give the estimate $y_t = 1.48872(5)$ for the 
XY universality class \cite{my3Nclock} as a black square and at 
$\overline{U}_C=-0.17455125(500)$ the estimate $y_t =  1.58737472(29)$
for the Ising universality class \cite{CB_Ising_2024}.
The dashed line is the result of a fit with the Ansatz~(\ref{yfit}). 
Details are discussed in the text.
}
\end{center}
\end{figure}

In Fig. \ref{yteffplot} we plot the effective $y_{t,eff}$ obtained from the
slopes of $Z_a/Z_p$, $U_4$, and $\xi_{2nd}/L$ at $\beta_f$ as a function of
$\bar{U}_C=[\bar{U}_C(\lambda,\mu,L_{max})+\bar{U}_C(\lambda,\mu,L_{min})]/2$.

Our numerical estimates obtained from the slopes of the three different 
quantities agree reasonably well. The estimates of $y_{t,eff}$ 
interpolate smoothly 
between the values for the XY and the Ising universality class.
We have fitted the data obtained for the slope of $Z_a/Z_p$ with the 
Ansatz 
\begin{equation}
\label{yfit}
y_{t,eff} = y_{XY} +c_2 \overline{U}_C^2 + c_2 \overline{U}_C^4 \;.
\end{equation}
We get $y_{t,XY} =1.48860(11)$, $c_2 =2.588(40)$, and $c_4=20.4(1.9)$.
This is given as dashed line in the plot.

We abstain from computing $\eta_{eff}$, since the values of the Ising and 
the XY universality class differ only by little.

\section{First order phase transitions}
\label{FirstOrder}
For $(\lambda,\mu)$ on the upper left of the decoupled Ising line, the RG flow 
runs off and a fluctuation induced first order phase transition is expected.
Since the RG flow is slow in the neighborhood of the decoupled Ising fixed 
point, there is a large range of parameters, where 
the first order transition is (very) weak. Hence in Monte Carlo simulations
of lattice models or in experiments, the first order nature of the 
transition can not be directly detected in this range of parameters. 

We simulate at several values of $(\lambda,\mu)$ close to the improved line.
We demonstrate the first order nature of the transition
for $(\lambda,\mu)$, where the first order transition is rather
strong. At these values of $(\lambda,\mu)$, we set the scale by computing 
the second moment correlation length at the transition temperature in the
high temperature phase $\xi_{2nd,high}$.  

In the following we first discuss the study of the strong first order 
transitions, sec. \ref{strong_transition}. Then the RG flow to stronger 
and stronger transitions is demonstrated, including the weak region, 
sec. \ref{weak_transition}. By using the RG flow, we obtain estimates
of $\xi_{2nd,high}$ at the transition for $(\lambda,\mu)$ in the weak 
transition range.

\subsection{Strong first order transitions}
\label{strong_transition}
We use the same simulation program as for the second order part of the 
$(\lambda,\mu)$ plane. To keep the study simple, we abstain from using methods
specially designed to study first order transitions such as the multicanonical
\cite{BeNe91,BeNe92,Ja98} or the Wang-Landau \cite{WaLa01,WaLa01E} method.

At the transitions temperature, each phase has the same weight 
\cite{BoKo90,BoJa92}. Here we have the disordered phase and the fourfold
degenerate ordered phase. Hence, 
$Z_a/Z_p$, $U_4$, and $U_C$  assume the values $1/5$, $5/4$, and   $-5/16$
at the transition temperature, in the limit $L \rightarrow \infty$.
For finite $L$, we take the value of $\beta$, where $Z_a/Z_p=1/5$ is assumed,
as an estimate of the transition temperature.

\begin{figure}
\begin{center} 
\includegraphics[width=14.5cm]{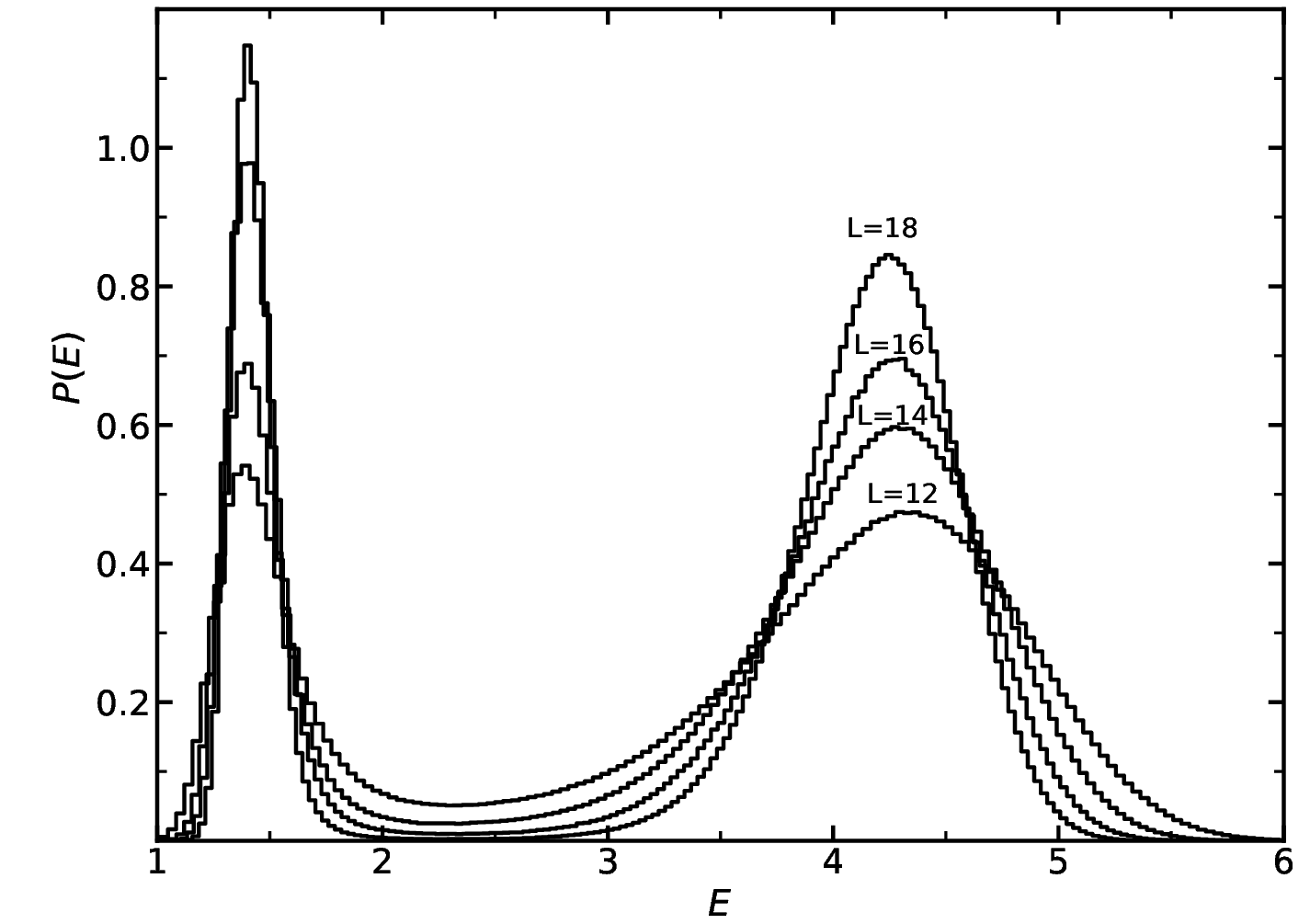}
\caption{\label{Histo}
We give the histograms of the energy density at
$(\lambda,\mu)=(0.65,2.1)$
for the linear lattice sizes $L=12$, $14$, $16$, 
and $18$. The data are reweighted to the Boltzmann distribution 
for $\beta=0.27001$, which is our
estimate of the inverse of the transition temperature.
The probability density at the minimum between to peaks is 
$P_{min} \approx 0.051$, $0.025$, $0.010$, and $0.0026$, for $L=12$, $14$, 
$16$ and $18$, respectively.
}
\end{center}
\end{figure}

First we simulated at values of $(\lambda,\mu)$ with strong first 
order transitions, such that the first order nature of the transition
can be demonstrated clearly. After a few preliminary simulations, 
we focus on $(\lambda,\mu)=(0.65,2.1)$, $(0.65,2.0)$, and $(0.675,2.1)$.
We performed simulations without pre-binning, storing the values
of a few observables computed for individual configurations. In particular, 
we have stored the energy, Eq.~(\ref{energy}). 
Plotting the estimates of the observables versus the Monte Carlo time, 
we see that with increasing lattice size metastabilities arise.  
We go up to linear lattice sizes, where we still get several transitions
between the phases in a simulation that takes a few days. 
In Fig. \ref{Histo} we plot the histogram of the 
energy density, reweighted to the estimate of the transition temperature
for $(\lambda,\mu)=(0.65,2.1)$  for the lattice sizes $L=12$, $14$, $16$, 
and $18$. We can see a clear double peak structure. The peaks become sharper
with increasing lattice size. The probability density at the minimum 
between the peaks rapidly decreases with increasing lattice size.
This behavior can also be seen in the histograms of other observables.
Similar observations hold for $(\lambda,\mu)=(0.65,2.0)$, and $(0.675,2.1)$.
The first order nature of the transition is clearly demonstrated. 

Based on the histogram of the observable $A$ an estimate of the transition 
temperature is obtained in the following way: First we select a value 
$A_{cut}$ between the peaks. Let us assume that $A$ is larger for ordered
configurations than for disordered ones. Then the estimate of the 
transition temperature is obtained by solving 
\begin{equation}
\int_{A_{min}}^{A_{cut}} \mbox{d} A \;P_{\beta}(A) =\frac{1}{n_{order}}
\int_{A_{cut}}^{A_{max}} \mbox{d} A \; P_{\beta}(A)
\end{equation}
with respect to $\beta$, where $n_{order}$ is the degeneracy of the 
ordered phase. We simulate at a value $\beta_s$ of the inverse 
temperature, close to $\beta_t$. The histogram is obtained as a function 
of $\beta$ in the neighborhood of $\beta_s$ by using reweighting.
The choice of $A_{cut}$ is ad hoc to some extend. However the dependence of
the estimate of $\beta_t$ on $A_{cut}$ rapidly decreases with increasing 
$L$, since $P(A)$ between the peaks rapidly decreases. For example for 
$(\lambda,\mu)=(0.65,2.1)$ we get from the histogram of the energy density
$\beta_t=0.270065(3), 0.270031(4)$, $0.270036(8)$, and $0.270022(8)$ for 
$L=12$, $14$, $16$, and $18$, respectively.
The value of $E_{cut}$ is roughly taken as the minimum of $P(E)$ between
the peaks. Using different simulations, with pre-binning and higher 
statistics, requiring that $Z_a/Z_p=1/5$, we obtain
$\beta_t =0.269994(4), 0.270003(4)$, $0.270007(4)$, and $0.270011(4)$, 
for $L=12$, $14$, $16$, and $18$, respectively.  It is plausible that
these two estimates converge to a unique value in the limit 
$L \rightarrow \infty$. 
As our final estimate we take $\beta_t=0.27001(1)$ obtained by 
requiring $Z_a/Z_p=1/5$. The error bar takes into account the variation 
of the estimate of $\beta_t$  with the linear lattice size $L$. Our 
estimates of $\beta_t$ for $(\lambda,\mu)$ in the strong first order
range are summarized in table \ref{table_beta_t}.
In order to quantify the strength of the transition, we compute the 
second moment correlation length $\xi_{2nd}$, eq.~(\ref{xi2nd}).  
To this end, we simulated at two values of $\beta$ close to the value of 
$\beta_t$ given in table \ref{table_beta_t}.
We started the simulations with the configuration $\vec{\phi_x} =(0,0)$ for 
all sites $x$. In the simulation, this configuration evolves into the 
disordered phase. The linear lattice
size is chosen such that the tunneling time to an ordered phase is very large 
compared with the simulation time, which is a few days on a single core.   
We monitored the evolution of the observables to ensure that we stay in the 
disordered phase. For this ensemble we compute the second moment correlation 
length as defined in eq.~(\ref{xi2nd}). Then we interpolate $\xi_{2nd}$ to 
$\beta_t$ of table \ref{table_beta_t}.  The results are given in the 
fourth column of  table \ref{table_beta_t}.

Below in Sec. \ref{universal_ratios} we compute 
ratios of quantities in the disordered and in one of the ordered phases 
at the transition temperature.

\begin{figure}
\begin{center}
\includegraphics[width=14.5cm]{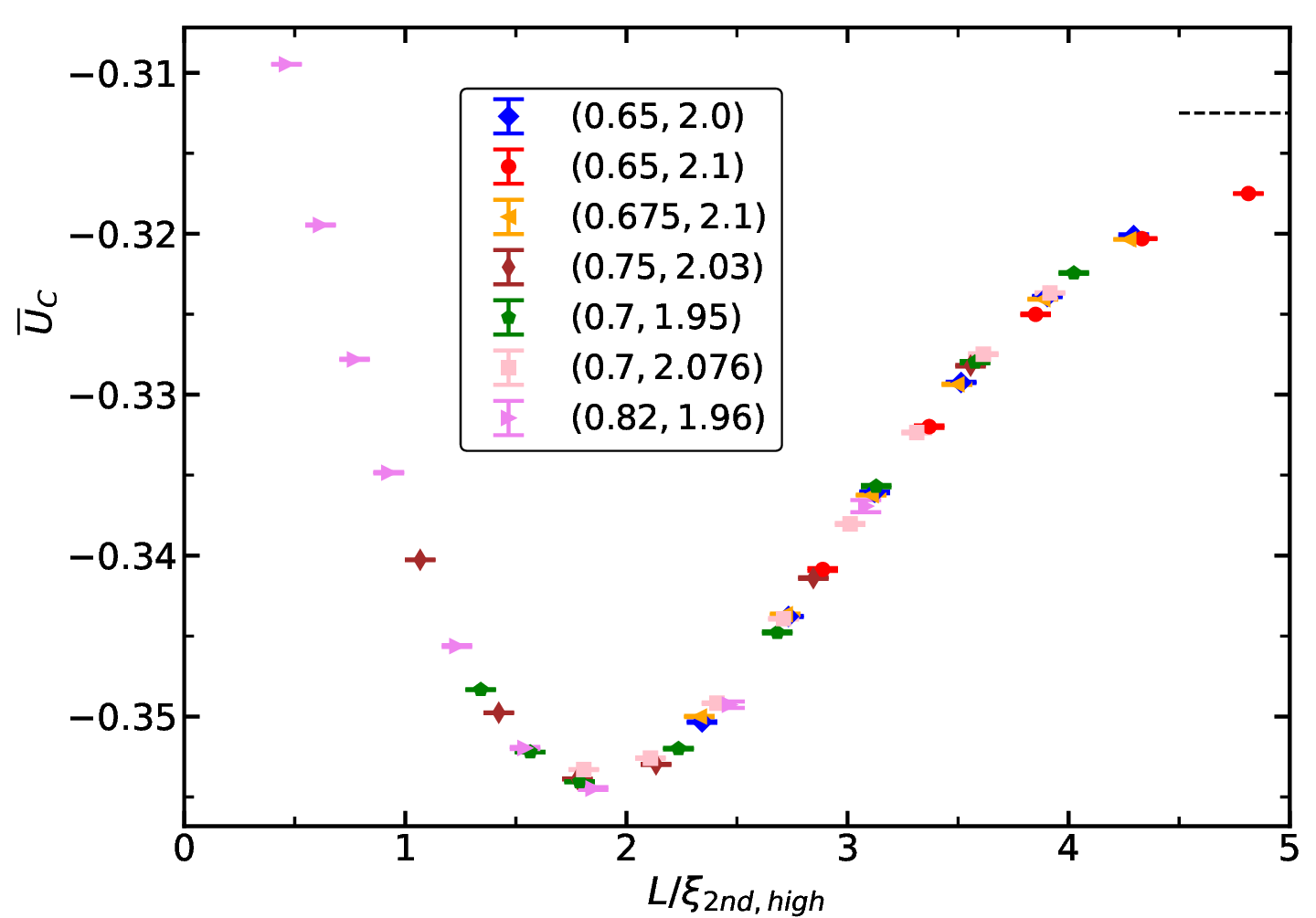}
\caption{\label{col0}
We plot $U_C$ at $Z_a/Z_p=1/5$ versus $L/\xi_{2nd,high}$ for the 
values of $(\lambda,\mu)$ given in the caption. The dashed line on the 
right indicates the value $U_C = -5/16$ assumed in the limit 
$L \rightarrow \infty$.
}
\end{center}
\end{figure}

\begin{table}
\caption{\sl \label{table_beta_t}
Estimates of $\beta_t$ for values of $(\lambda,\mu)$ with a strong first order 
transition. 
For $\xi_{2nd,high}$ we give in $()$ the statistical error at 
the given value of $\beta_t$, while in $[]$ the uncertainty induced by 
the error  of $\beta_t$ is given.
For a discussion see the text.
}
\begin{center}
\begin{tabular}{llll}
\hline
\mc{1}{c}{$\lambda$} & \mc{1}{c}{$\mu$}  &  \mc{1}{c}{$\beta_t $} & 
\mc{1}{c}{$\xi_{2nd,high}$} \\
\hline
0.65      & 2.1    & 0.27001(1) & \phantom{0}4.165(13)[11] \\
0.65      & 2.0    & 0.297715(10) & \phantom{0}5.123(5)[12]  \\
0.675     & 2.1    & 0.28735(1)  & \phantom{0}5.150(7)[11] \\
0.7       & 2.076  & 0.307766(5) & \phantom{0}6.640(17)[20] \\
0.7       & 1.95   & 0.335044(5) & \phantom{0}8.947(20)[27] \\
0.75      & 2.03   & 0.339872(6) & 11.246(12)[61]\\
0.82      & 1.96   & 0.373841(5) & 25.95(12)[46] \\
\hline
\end{tabular}
\end{center}
\end{table}
Next we study the scaling of $\overline{U}_C$ with $L/\xi_{2nd,high}$. 
Here we determine $\beta_f$ by requiring that $Z_a/Z_p$ assumes the 
value $(Z_a/Z_p)_f=1/5$. In Fig. \ref{col0} we plot $\overline{U}_C$
versus $L/\xi_{2nd,high}$.
We give results for all values of $(\lambda,\mu)$, where we have 
directly computed $\xi_{2nd,high}$.  We observe that for different 
$(\lambda,\mu)$ the estimate of $\overline{U}_C$ versus $L/\xi_{2nd,high}$
fall approximately on a unique scaling curve. The limit 
$L/\xi_{2nd,high} \rightarrow \infty$ is approached non-monotonically.
We have redone the analysis by fixing $Z_a/Z_p$ to its decoupled Ising 
value. We obtain qualitatively the same result.

\subsection{RG flow towards strong first order transitions}
\label{weak_transition}
Here we demonstrate the RG flow towards stronger and stronger first 
order transitions.  The RG flow connects weak transitions with the 
strong ones. This allows us to quantify the strength of the transition
also in the weak region, where the first order nature of the transition 
can not be detected directly. To this end, we study the slow flow that is 
characterized by the finite size behavior of $\overline{U}_C$.

\begin{table}
\caption{\sl \label{table_xi_extra}
Estimates of the matching factor $c_{1,2}$, where $(\lambda_1,\mu_1)$ 
is given by the $(\lambda,\mu)$ in the same row, and $(\lambda_2,\mu_2)$
by $(\lambda,\mu)$ in row one above. In the last column we give an estimate of
$\xi_{2nd,high}$. The starting point is $(\lambda,\mu)=(0.82,1.96)$, 
where we have directly computed $\xi_{2nd,high}$. Then for 
$(\lambda,\mu)=(0.8,1.8)$ we get 
$\xi_{2nd,high}=25.95(12)[46]/0.660(3)=39.3(1.1)$ and so on.
For a discussion see the text.
}
\begin{center}
\begin{tabular}{lllc}
\hline
\mc{1}{c}{$\lambda$}& \mc{1}{c}{$\mu$} &\mc{1}{c}{ $c_{1,2}$} & $\xi_{2nd,high}$\\
\hline    
0.82      & 1.96   &      &  25.95(12)[46]   \\
0.8       &  1.8   & 0.660(3) & 39.3(1.1) \\
0.9       & 1.89   & 0.485(4) & 81.1(2.9) \\
0.9       & 1.7    & 0.304(6) & 267.(15.) \\
1.0       & 1.8    & 0.447(5) & 597.(41.) \\
1.03      & 1.77   & 0.431(6) & 1384.(117.) \\
1.06      & 1.75   & 0.414(6) &  3343.(335.) \\
1.1       & 1.72   & 0.21(1)  &  15921.(2471.) \\
1.12      & 1.7    & 0.330(6) &  48246.(8520.) \\
\hline
\end{tabular}
\end{center}
\end{table}

For two values $(\lambda_1,\mu_1)$ and $(\lambda_2,\mu_2)$ we
determine a scale factor $c_{1,2}$ by requiring that
\begin{equation}
\label{matching_condition}
\overline{U}_{C,1}(L) =  \overline{U}_{C,2}(c_{1,2} L) \;,  
\end{equation}
where the second subscript indicates the value of $(\lambda,\mu)$.
This equation is solved numerically for each linear lattice size  $L$
and each value of $(\lambda,\mu)$ simulated in the first order range.
The first requirement is that  $(\lambda_1,\mu_1)$ and $(\lambda_2,\mu_2)$
have overlapping ranges of $\overline{U}_{C}$ for the linear lattice sizes 
simulated. Furthermore, for simplicity, we only consider a range of 
lattice sizes, where $\overline{U}_{C}$ is monotonically decreasing with 
increasing lattice size $L$. 
In a first step, we determine two lattice sizes $L_{1}$ and $L_{2}$ for 
$(\lambda_2,\mu_2)$ such that $L_1$ is the largest lattice size simulated with 
$\overline{U}_{C,2}(L_1) \ge \overline{U}_{C,1}(L)$ and $L_2$ is the smallest 
linear  lattice size simulated  such that 
$\overline{U}_{C,2}(L_2) \le \overline{U}_{C,1}(L)$.
If such a pair of lattice sizes had been simulated, we interpolate
$\overline{U}_{C,2}$ linearly in the logarithm of the linear lattice size.

Let us go through the steps at the example of 
$(\lambda_1,\mu_1) = (1.03,1.77)$ and $(\lambda_2,\mu_2) = (1.0,1,8)$
fixing $Z_a/Z_p=1/5$. 
For $(\lambda_1,\mu_1) = (1.03,1.77)$, we get $\overline{U}_C=-0.235186(22)$,
$-0.236734(19)$, $-0.238049(26)$, $-0.240384(29)$, $-0.242533(35)$, 
$-0.245945(44)$, $-0.248824(51)$, $-0.251419$, and $-0.255760(81)$ 
 for $L=12$, $14$, $16$, $20$, $24$, $32$, $40$, $48$, and $64$, 
respectively.
For $(\lambda_2,\mu_2) = (1.0,1,8)$ we get 
$\overline{U}_C=-0.247992(25)$, $-0.253604(37)$, $-0.258107(50)$, $-0.265192(56)$
$-0.270875(86)$, $-0.27949(14)$, and $-0.28671(34)$ for $L=16$, $24$, $32$, 
$48$, $64$, $96$, and $128$, respectively.

For example $\overline{U}_C=-0.248824(51)$ for $L=40$ and 
$(\lambda_1,\mu_1) = (1.03,1.77)$ is encapsulated by 
$\overline{U}_C=-0.247992(25)$ and $-0.253604(37)$ for $L_1=16$ and $L_2=24$
and $(\lambda_2,\mu_2) = (1.0,1,8)$. Linearly interpolating in $\log L$
we find that $\overline{U}_C=-0.248824(51)$ is assumed for $L=17.00(7)$. 
Hence $c_{1,2}=0.425(2)$.  Starting from $L=48$ and $64$ for 
$(\lambda_1,\mu_1)$ we arrive at the estimates $c_{1,2}=0.4274(21)$ and 
$0.4307(24)$. As our final estimate we take $c_{1,2}=0.431(6)$. The error
bar takes into account the variation of the estimate with the linear lattice 
size.  Our results for all $(\lambda,\mu)$ with a weak first order 
transition are summarized in table \ref{table_xi_extra}. We start with 
$(\lambda,\mu)=(0.82,1.96)$, which is the value with the largest
$\xi_{2nd,high}$ that we have directly determined. We have computed 
estimates of $\xi_{2nd,high}$ by using 
$\xi_{2nd,high,1} = \xi_{2nd,high,2}/c_{1,2}$ recursively. 
For simplicity, we have 
added up the errors due to the uncertainty of $c_{1,2}$. Note that this
procedure is inspired by ref. \cite{Cara95}.
In Fig. \ref{col0}, as a check, we plot $\overline{U}_C$ versus 
$L/\xi_{2nd,high}$ for the values of $(\lambda,\mu)$ given in table
\ref{table_xi_extra}. Indeed we see a nice collapse on a unique line.

\begin{figure}
\begin{center}
\includegraphics[width=14.5cm]{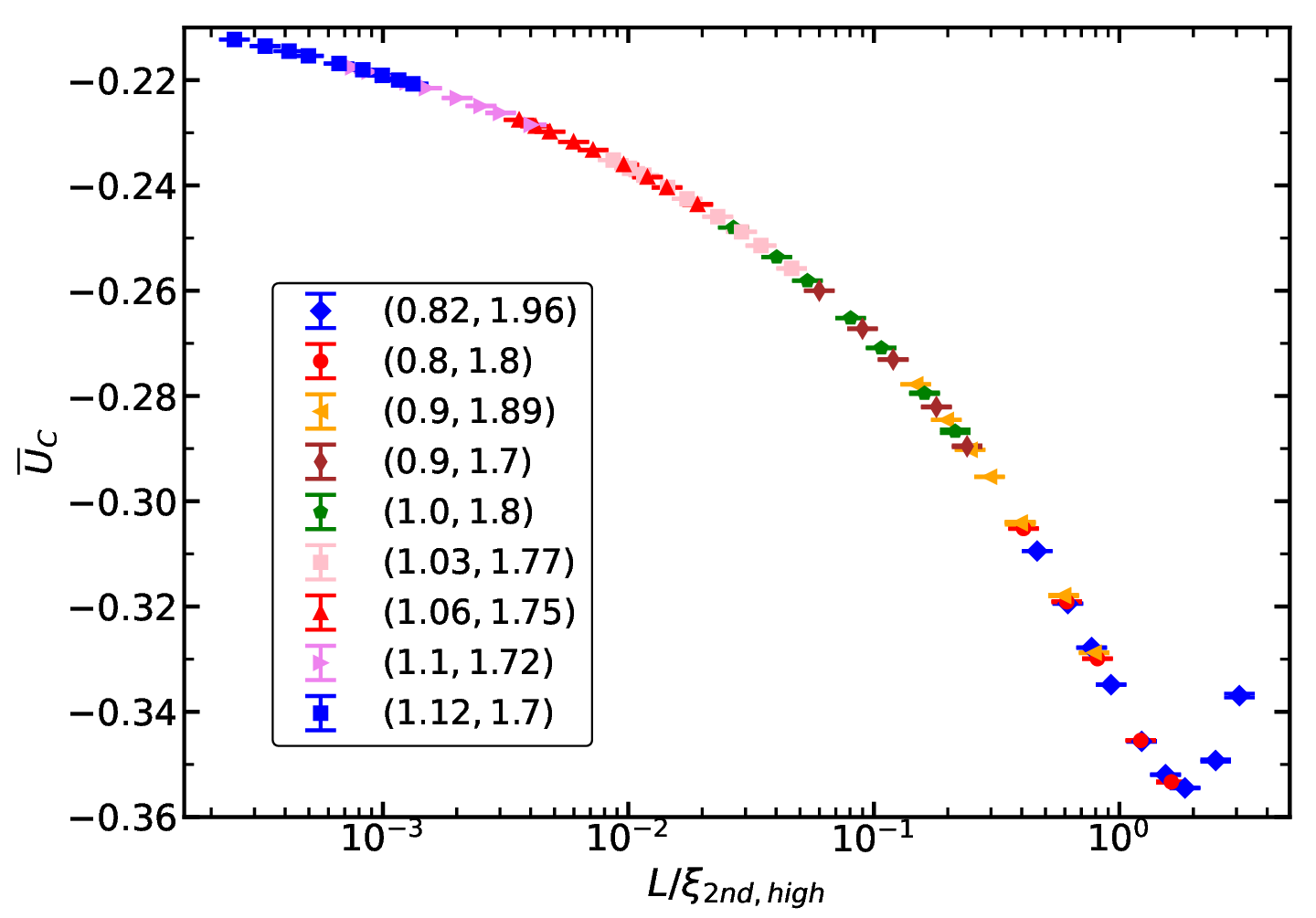}
\caption{\label{col1}
Similar to Fig. \ref{col0} but for $(\lambda,\mu)$ with weak first order
transitions.
We plot $U_C$ at $Z_a/Z_p=1/5$ versus $L/\xi_{2nd,high}$ for the
values of $(\lambda,\mu)$ given in the caption. We take $\xi_{2nd,high}$
from table \ref{table_xi_extra}.
}
\end{center}
\end{figure}

\subsubsection{RG flow in the neighborhood of the DI fixed point}
Here we study $\tilde u$, defined in eq.~(\ref{tildeU}), which is a function
of 
\begin{equation}
\tilde{U}_{C} = \overline{U}_{C} - \overline{U}_{C,DI}^*  \;. 
\end{equation}
Below we define $\overline{U}_{C}$ by 
requiring $Z_a/Z_p=(Z_a/Z_p)_{DI}^* = (Z_a/Z_p)_{Ising}^{*,2}$. The numerical 
method to compute $\tilde u$ is the same as for $u$ in section 
\ref{slowflow_theory}.

\begin{figure}
\begin{center}
\includegraphics[width=14.5cm]{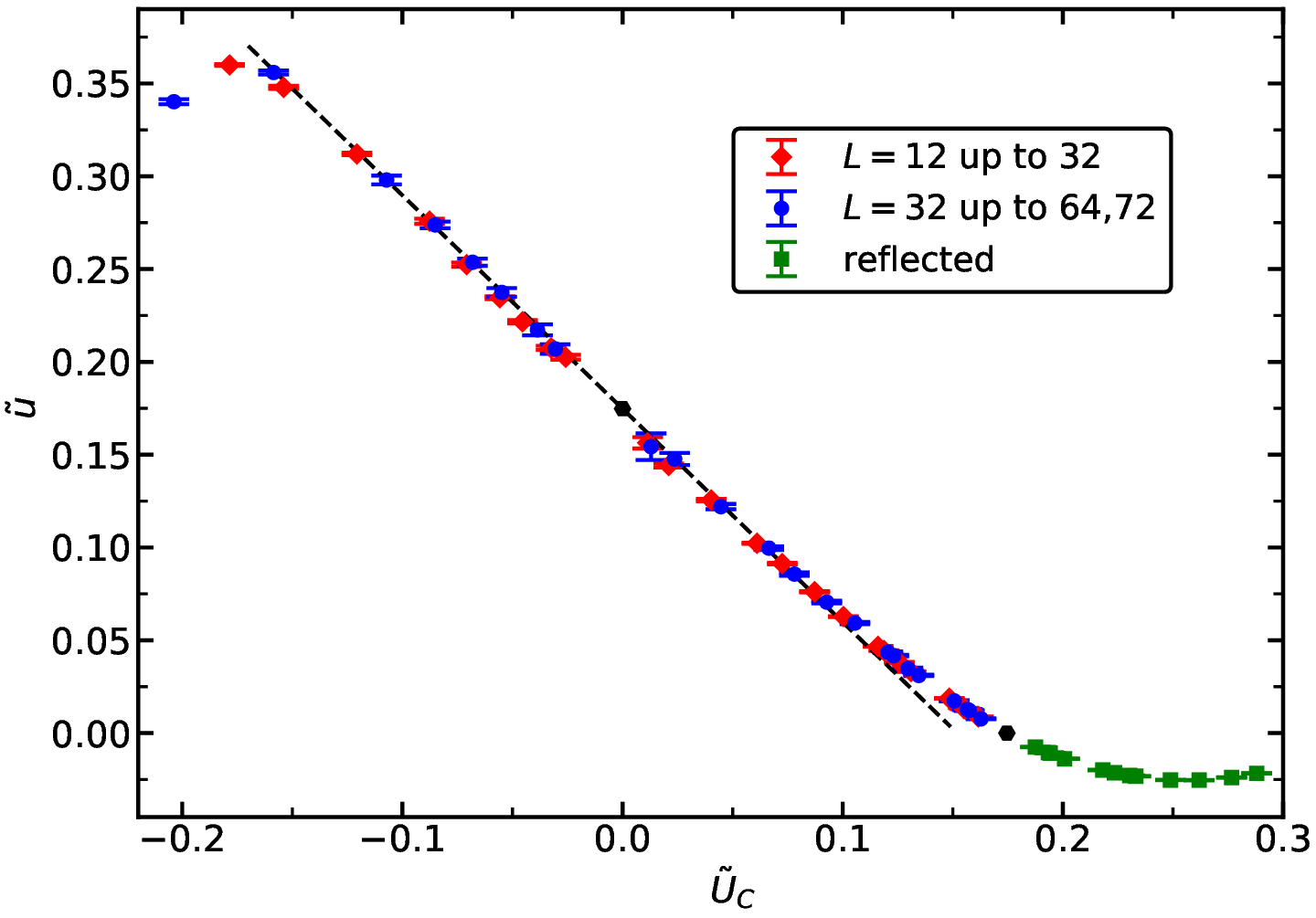}
\caption{\label{usub_plot}
We plot $\tilde u(\tilde U_C)$ versus $\tilde U_C$ for the two
fit ranges $L=12$ up to $L=32$ and $L=32$ up to $L=64$ or $72$.
The hexagons give the known estimates for the fixed points. The data 
denoted by reflected, are the ones for the fit range $L=32$ up to $L=64$ 
or $72$ reused, exploiting that 
$\overline{U}_C(L,\lambda,\mu)=-\overline{U}_C(L,\lambda,-\mu)$.
The dashed line gives the linear approximation of $\tilde u$ with $a_1=-1.15$.
For a discussion see the text.
}
\end{center}
\end{figure}

We consider values of $(\lambda,\mu)$ close to the improved line. We determine
$\tilde u$ by fitting our data with the Ansatz
\begin{equation}
\tilde U_C = a L^{\tilde u}
\end{equation}
for the two ranges of lattice sizes $L=12$ up to $L=32$ and $L=32$ up to $L=64$.
These estimates are plotted in FIG. \ref{usub_plot}.

We analyze the numerical results for $\tilde u$  by using Ans\"atze of
the form
\begin{equation}
\tilde u = a_0 + a_1 \tilde U_C + a_2  \tilde U_C^2 + ... \; .
\end{equation}
Note that here, in contrast to eqs.~(\ref{ufit2},\ref{ufit4}) odd powers appear.
We performed fits for the two choices of the fit range: Either $L_{min}=12$
and $L_{max}=32$ or $L_{min}=32$
and $L_{max}=64$, or $72$.
In order to check for systematic errors we varied the range of $(\lambda,\mu)$
taken into account in the fits and the maximal power in the Ansatz. 
We find that the estimates of $a_0$ are
compatible with the theoretic prediction for $y_{DI}$. In the case of $a_1$
we get similar results for different fits, while for higher order coefficient 
the results start to scatter. As final result we take $a_1=-1.15(3)$.
In FIG. \ref{usub_plot} we see that the linear approximation nicely
reproduces the data for a large neighborhood of $\tilde U_C=0$.

Finally we check, whether $\tilde u= y_{DI} - 1.15 \tilde U_C$ is consistent
with the factors $c_{1,2}$ given in table \ref{table_xi_extra}.
To this end we numerically integrate $\tilde U_C$ using 
$\tilde u= y_{DI} - 1.15 \tilde U_C$. We start the integration 
at $\tilde U_C$ obtained for $L=64$ at $(\lambda_1,\mu_1)$ and stop
the integration when $\tilde U_C$ obtained for $L=64$ at $(\lambda_2,\mu_2)$
is reached. Then the exponential of minus the interval of the integration
gives the estimate of $c_{1,2}$. We find that the results are fully consistent
for
$(\lambda_1,\mu_1)=(0.9,1.89)$ and $(\lambda_2,\mu_2)=(0.9,1.7)$ and down
in table \ref{table_xi_extra}.

\subsection{Universal ratios at the first order transition}
\label{universal_ratios}
Here we follow Refs. \cite{Arnoldetal97,ArZh97,ArYa97,ArZh97e}. 
In the limit of 
weaker and weaker first order transitions, as the decoupled Ising fixed 
point is approached, ratios of quantities computed
in one of the ordered and the disordered phase assume universal values.
In Refs. \cite{Arnoldetal97,ArYa97,ArZh97e,ArZh97} such ratios are computed  
for the correlation length, the magnetic susceptibility and the specific heat. 
In Ref. \cite{ArZh97} Monte Carlo simulations of the Ashkin-Teller model
on the simple cubic lattice were performed. In Refs. 
\cite{Arnoldetal97,ArYa97,ArZh97e} 
the universal ratios were computed by using the $\epsilon$-expansion.
Numerical results are compared in table I of Ref. \cite{Arnoldetal97}.
As their extrapolated Monte Carlo result they quote $C_+/C_-=0.069(8)$,
$\xi_+/\xi_-=1.7(2)$, and $\chi_+/\chi_-=4.1(5)$. Here the subscripts
$+$ and $-$ indicate the high and low temperature phases, corresponding 
to disorder and order. $C$, $\xi$, and $\chi$ denote the specific heat, 
the second moment correlation length and the magnetic susceptibility.
In Refs. \cite{ArYa97,ArZh97e} the authors have
computed the leading order (LO) and next-to-leading order (NLO)  result
in the $\epsilon$-expansion for  $C_+/C_-$ and $\xi_+/\xi_-$,
while for $\chi_+/\chi_-$ they have the next-to-next-to-leading order (NNLO) 
in addition. The authors did 
not perform a resummation of the series since there are contribution that
are logarithmic in $\epsilon$. The LO and NLO results differ a lot. No precise 
estimates for $d=3$ are obtained. For a more detailed discussion see
Ref. \cite{Arnoldetal97}. 

Here we have simulated at the values $(\lambda,\mu)=(0.65,2.1)$, 
$(0.65,2.0)$, $(0.675,2.1)$, $(0.7,2.076)$, $(0.7,1.95)$, and 
$(0.75,2.03)$. The estimates of $\beta_t$ are taken from table
\ref{table_beta_t} above. We realized that
$(\lambda,\mu)=(0.82,1.96)$ is too ambitious here.
Instead, we have added one values of $(\lambda,\mu)$ with a strong
first order transition. Analysing the behavior of $Z_a/Z_p$, we get 
$\beta_t=0.235375(5)$ at $(\lambda,\mu)=(0.6,2.07)$.

To get into an ordered phase, we started the simulations with 
$\phi_{x,0}=\phi_{x,1} = c$ for all sites $x$. Throughout we took $c=1$.
After some preliminary simulations, we decided 
to take $L \gtrapprox 25 \xi_{2nd,-}$.  This way we stay in the initial
phase and finite size effects are negligible.
Computing $\chi$, $\xi_{2nd}$, and $\xi_{exp}$, in the low temperature 
phase, we follow Ref. \cite{MH_ratio_10}. In total we have used about
one year of CPU time on a single
core of an Intel(R) Xeon(R) CPU E3-1225 v3 running at 3.20 GHz for 
the simulations in the ordered phase.

\begin{figure}
\begin{center}
\includegraphics[width=14.5cm]{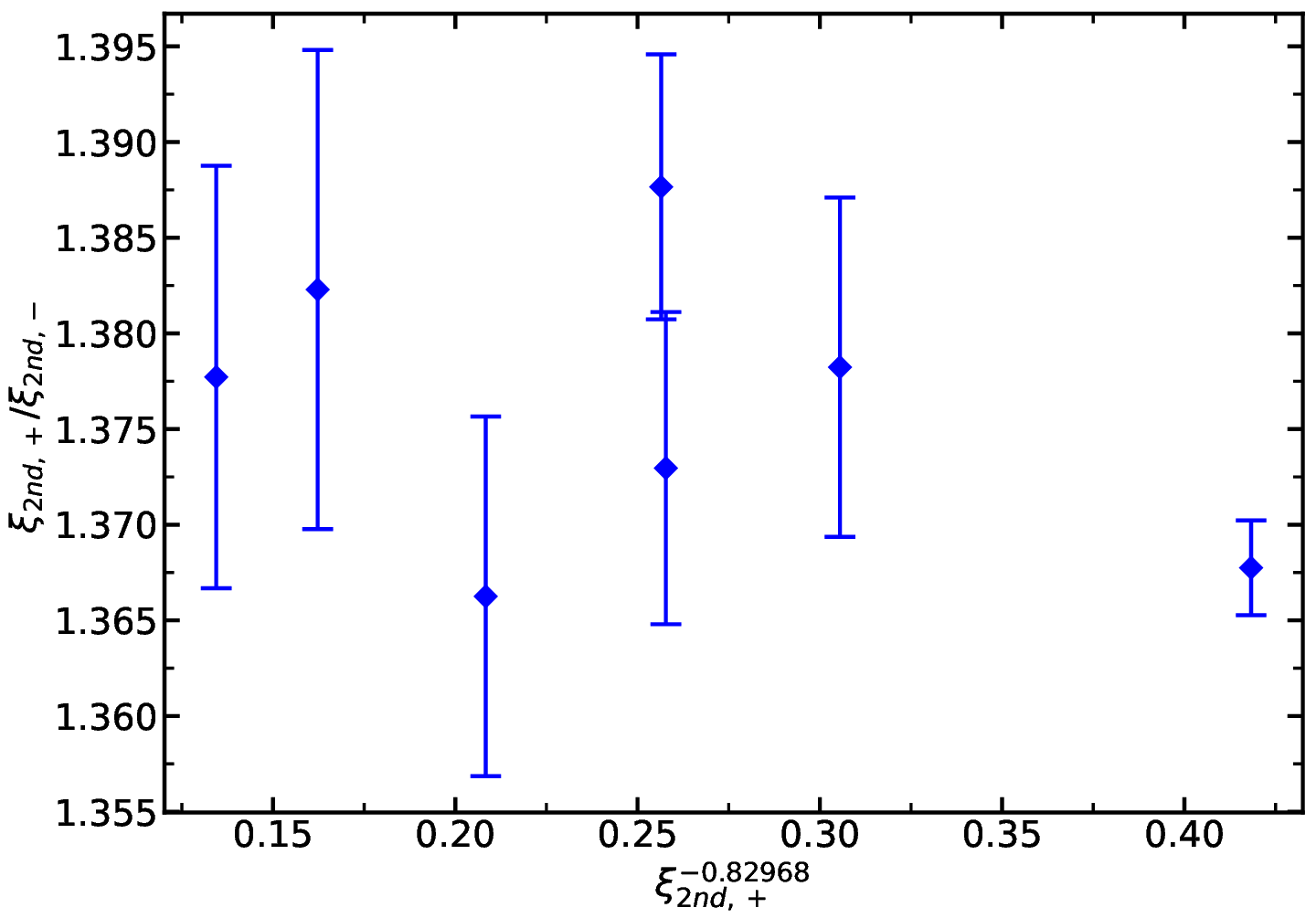}
\caption{\label{ratio_xi2nd}
We plot the ratio $\xi_{2nd,+}/\xi_{2nd,-}$ versus $\xi_{2nd,+}^{-\omega_I}$.
For a discussion see the text.
}
\end{center}
\end{figure}

In Fig. \ref{ratio_xi2nd} we plot our results for the ratio 
$\xi_{2nd,+}/\xi_{2nd,-}$ versus $\xi_{2nd,+}^{-\omega_I}$, following
Ref. \cite{ArZh97}. Here we give the statistical error for the 
simulation at the given value of $\beta_t$. We checked only in two 
cases the dependence of the ratios on $\beta_t$. The error due to 
the uncertainty of $\beta_t$ is of similar size as the statistical one 
at given $\beta_t$. 
Within the statistical error, the ratio $\xi_{2nd,+}/\xi_{2nd,-}$ remains 
almost constant with increasing $\xi_{2nd,+}$. We quote 
\begin{equation}
\frac{\xi_{2nd,+}}{\xi_{2nd,-}} =1.38(3) 
\end{equation}
as result for the scaling limit,  which is smaller than the estimate of 
Ref. \cite{ArZh97}.
Looking at FIGs. 10 and 11 of Ref. \cite{ArZh97}, one might suspect that
in the range that is taken into account for the extrapolation,  
not only leading corrections but also subleading corrections
are not negligible. This might explain the deviation.

In the case of $\chi_+/\chi_-$ estimates do scatter slightly.
For the values $(\lambda,\mu)=(0.65,2.0)$ and $(0.675,2.1)$ we have 
approximately the same correlation length $\xi_{2nd,+}$  at the transition.
We get $\chi_+/\chi_-=3.874(10)$ and $3.971(9)$ for 
$(\lambda,\mu)=(0.65,2.0)$ and $(0.675,2.1)$, respectively.  This might
be due to the fact that $(\lambda,\mu)=(0.65,2.0)$ and $(0.675,2.1)$
are on different sides of the improved line.  Overall there is little 
dependence $\chi_+/\chi_-$ on $\xi_{2nd,+}$. As our estimate for the
scaling limit we quote
\begin{equation}
\frac{\chi_+}{\chi_-} = 3.9(1)  \; ,
\end{equation}
which is consistent with the estimate of Ref. \cite{ArZh97}.

In the case of $C_+/C_-$ the estimates increase with increasing 
$\xi_{2nd,+}$. For $(\lambda,\mu)=(0.6,2.07)$ we get $C_+/C_-=0.05898(9)$
while for $(\lambda,\mu)=(0.75,2.03)$ we get $0.0700(7)$. Here we 
abstain from quoting a scaling limit. The values of $C_+/C_-$ are in a
range that is compatible with the estimate $C_+/C_-=0.069(8)$ of
Ref. \cite{ArZh97}.

\section{Summary and outlook}
\label{summary}
We study the RG flow in the three-dimensional two component $\phi^4$-model
with cubic symmetry, Eq.~(\ref{Hamiltonian}). The problem, for a general 
number $N$ of components of the field, has been studied
by using field theoretic methods starting from the early days of the RG. 
For recent work see  Ref. \cite{AharonyNeu} and references therein.

The purpose of the present work is to advance the methodology of using
Monte Carlo simulations of lattice models in combination with finite size 
scaling of dimensionless quantities. Furthermore we provide numerical 
results for the RG flow that could be compared with those
obtained for other models as for example the Ashkin-Teller model.

In order to get accurate results from relatively small lattices, 
we first determine in Sec. \ref{slowflownum} the line of the slow flow in the 
$(\lambda,\mu)$ plane. This line is a generalization of improved models.
The one-dimensional slow RG flow is characterized by a modified 
$\beta$-function $u(\overline{U}_C)$, Eq.~(\ref{flowequation}), where 
$\overline{U}_C$ is $U_C$, Eq.~(\ref{UCdef}), on the critical surface. 
For the precise 
definition see Sec. \ref{slowflow_theory}. We obtain accurate numerical
results for the whole range from the decoupled Ising fixed point to 
the $O(2)$ invariant one. In particular
$\lim_{\overline{U}_C \rightarrow 0} u(\overline{U}_C) =Y_4$,
where $Y_4$ is the RG exponent of the cubic perturbation at the 
$O(N)$ invariant fixed point. Below, in Sec. \ref{Y4results},
we compare our accurate result with ones given in the literature.
Following Ref. \cite{AharonyNeu} we compute in Sec. \ref{yteff} 
an effective thermal RG exponent as a function of $\overline{U}_C$, see
Eq.~(\ref{yfit}).  In the limit $\overline{U}_C \rightarrow 0$ we 
get $y_{t,XY}=1.48860(11)$, which is compatible with our recent estimate
$y_{t,XY}=1.48872(5)$ \cite{my3Nclock} and  $y_{t,XY}= 1.48864(22)$ 
\cite{che19} obtained by using the CB method.

In the second part of the numerical analysis, Sec. \ref{FirstOrder}, 
we consider
first order transitions. First we searched for values of $(\lambda,\mu)$,
where the transition is strong, such that the first order nature
of the transition can be easily established in the simulation.
Following 
refs. \cite{Arnoldetal97,ArYa97,ArZh97e,ArZh97} we have  determined
the ratios $\xi_+/\xi_-$,  $\chi_+/\chi_-$, and $C_+/C_-$ at
the first order transition, where the subscript indicates the phase
and $\xi$, $\chi$, and $C$ are the second moment correlation 
length, the magnetic susceptibility and the specific heat, 
respectively. At least for $\xi_+/\xi_-$ and $\chi_+/\chi_-$
we were able to extract the scaling limit in a straight forward way.
The estimates are more accurate than those of Ref. \cite{ArZh97},
where the authors simulated the Ashkin-Teller model.
Next we follow the RG flow from weak to strong first order
transitions. The RG flow is given by the 
modified $\beta$-function $\tilde u(\tilde{U}_{C})$,
where $\tilde{U}_{C} = \overline{U}_{C} - \overline{U}_{C,DI}^*$
and $\overline{U}_{C,DI}^*$ is the value at the decoupled Ising 
fixed point. It is well approximated by
$\tilde u = y_{DI} -1.15(3) \tilde{U}_{C}$ in a rather extended 
neighborhood of the decoupled Ising fixed point, where 
$y_{DI}$ is given in terms of Ising RG exponents, 
Eq.~(\ref{DIexponent}). 

\subsection{Estimates of $Y_4$ given in the literature}
\label{Y4results}
Our estimate $Y_4=u_0 =-0.1118(10)$  of Sec. \ref{slowflow_numerics}
can be compared with results obtained in the literature by using 
various methods. The estimates given in Ref. \cite{Carmona} obtained by
using the $\epsilon$-expansion and the expansion in three-dimensions fixed
are consistent with but less precise than our estimate. The same 
holds for the result obtained by using the functional renormalization 
group (FRG), Ref. \cite{Chle22}. The estimate of Ref. \cite{O2corrections}
is close to our result, however the difference is about twice as large as 
the sum of the two errors.
\begin{table}
\caption{\sl \label{Y4lit}
Estimates of $Y_4$ for the three-dimensional XY universality class given
in the literature obtained by various methods. The asterix on the error of
Ref. \cite{O2corrections} means  that, in contrast to other quantities 
determined by the authors, the error is estimated non-rigorously.
A detailed discussion is given in the text.
}
\begin{center}
\begin{tabular}{cccl}
\hline
  Ref.    &year   & method  &    \mc{1}{c}{$Y_4$}  \\
\hline
 \cite{Carmona}&2000 &$\epsilon$-exp. $5^{th}$ order& --0.114(4) \\
 \cite{Carmona}&2000 & 3D exp. $6^{th}$ order       & --0.103(8) \\
 \cite{O2corrections}&2020&CB                       & --0.11535(73$^*$) \\
 \cite{Chle22} &2022 &  FRG                         & --0.111(12) \\
 \cite{O234}   &2011 & MC, FSS                      & --0.108(6) \\
 \cite{Debasish}&2018& MC, FSS                      & --0.128(6) \\
 \cite{Shao20} &2020 & MC, FSS                      & --0.114(2) \\
 \cite{Cuomo24}&2024 & MC, FSS                      & --0.124(2) \\
  present work &2025 & MC, FSS                      & --0.1118(10) \\
\hline
\end{tabular}
\end{center}
\end{table}
Refs. \cite{O234,Debasish,Shao20,Cuomo24} perform Monte Carlo simulations
of lattice models as it is done in the present work. However the details 
differ significantly.

In Ref. \cite{O234} the improved $\phi^4$ model is simulated by using 
a hybrid of the single cluster algorithm and local Metropolis updates.
The exponents $Y_q$ are obtained from the finite size scaling behavior 
of the correlation functions $C_i$, see Eqs.~(16,17,18) of ref. \cite{O234}.
In particular for $q=4$:
\begin{equation}
C_4 = 
\sum_{abcd} \left \langle \sum_x Q_4^{abcd}(\vec{\phi}_x) Q_4^{abcd}(\vec{m}) 
\right \rangle \;,
\end{equation}
where $\vec{m} =\frac{\vec{M}}{|\vec{M}|}$ and $\vec{M}=\sum_x \vec{\phi}_x$. 
And $Q_4$ is the traceless symmetric combination of four instances of
the field that we use in section \ref{Model} and in eq.~(\ref{UCdef}) to define
the dimensionless quantity $U_C$.  
In addition 
\begin{equation}
D_4 = 
\frac{\sum_{abcd} 
\left \langle \sum_x Q_4^{abcd}(\vec{\phi}_x) Q_4^{abcd}(\vec{M}) \right \rangle}
{\langle \vec{M}^2 \rangle^{q/2} } \;,
\end{equation}
is used. Note that in $C_q$ the order parameter is normalized for each 
configuration, while in $D_q$ this is done for the expectation values.
One expects that both quantities have the same asymptotic behavior
$C_q, D_q \propto L^{Y_q}$ at criticality. However
the relative statistical error and corrections to scaling might differ.
The problem of both quantities is that with increasing $q$, 
the relative variance of $C_q$ and $D_q$
is increasing more and more with increasing linear lattice size $L$. 
Related to this fact,
in ref. \cite{O234}, in the case of the 2-component $\phi^4$ model, only 
linear lattice sizes up to $L=28$ were simulated.

In ref. \cite{Debasish} a completely different approach is taken.
The authors analyze the behavior of the two-point function
\begin{equation}
 C_q(r)  = \langle \exp(i q \theta_r) \exp(-i q \theta_0) \rangle
\sim \frac{a(Q)}{|r|^{2 D(q)} }
\end{equation}
in the three dimensional XY model on the simple cubic lattice at the 
critical temperature. The angle $\theta$ is defined by 
$\vec{s}_x =(\cos(\theta_x), \sin(\theta_x))$. Note that $D(q)=3-Y_q$
is the dimension of the field.
They simulated finite lattices. To cancel the finite size effects,
$r=(L/2,0,0)$ is used throughout.
They simulated the model by using the worm algorithm \cite{Prok01}.
Computing this correlation function in direct way, the relative variance 
increases strongly with increasing $q$.
They overcome the problem by using an iterative procedure, where 
ratios $C_{q+1}(r)/C_{q}(r)$ are computed. The authors extracted the differences
in the dimensions of the field by fitting the data using the Ansatz 
\begin{equation}
C_{q+1}(r)/C_{q}(r) = b r^{-2[D(q+1)-D(q)]}
\end{equation}
not taking into account corrections to scaling. The authors simulated 
lattices up to a linear size $L=120$.  
Their estimate of $Y_4$ deviates by more than twice the error from our
estimate. This might be due to the leading correction to scaling that is not
explicitly taken into account the analysis of the data. Furthermore the 
simulations are performed slightly off critical. The authors 
simulated at $\beta=0.4541652$, while the most recent estimates of the 
inverse critical temperature of the XY model on the simple cubic lattice are
$\beta_c=0.45416476(10)$, Ref. \cite{Deng19}, and $0.45416474(10)[7]$, Ref. 
\cite{myClock}.

The authors of Ref. \cite{Cuomo24} follow much the same approach as 
Ref. \cite{Debasish}. They discuss an improvement of the algorithm. 
As the authors of Ref. \cite{Debasish}, they simulate at $\beta=0.4541652$.
For small values of $q$, they perform simulations for linear 
lattice sizes up to $64$. In Fig 13 of Ref. \cite{Cuomo24}, 
the estimate of the scaling limit is obtained by a linear extrapolation 
in $L^{-1}$. Note however that 
the leading correction to scaling is proportional to $L^{-\omega}$.
Given the quoted error, the estimate of Ref. \cite{Cuomo24} is clearly 
inconsistent with ours.

The authors of Ref. \cite{Shao20}  study $\mathbb{Z}_q$ invariant 
models on the simple cubic lattice, where
$q=4$, $5$, and $6$. Their main focus is the effect of the dangerously
irrelevant perturbation in the low temperature phase.  They also 
provide results for the critical temperature that are closely related
to those obtained here. 
The authors of Ref. \cite{Shao20} simulated a model defined by the 
reduced Hamiltonian
\begin{equation}
 H =-\beta 
 \left[\sum_{<xy>} \cos(\theta_x-\theta_y)+h \sum_x \cos(q \theta_x) \right] \;,
\end{equation}
where $\theta_x \in [0,2 \pi) $. 
I.e. in contrast to our convention, $\beta$ multiplies the potential term.
In order to measure the unisotropy of the order parameter, they use the 
quantity
\begin{equation}
\label{phiq}
 \phi_q = \cos(q \Theta) \; ,
\end{equation}
where $\Theta=\arccos(M_0/M)$, $M_0=\sum_x \cos(\theta_x) $, 
$M_1=\sum_x \sin(\theta_x)$ and $M=(M_0^2+M_1^2)^{1/2}$. Here in our case 
$q=4$. 
We note that
\begin{equation}
\cos(4 \Theta)=\cos^4(\Theta)-6 \cos^2(\Theta) \sin^2(\Theta)+\sin^4(\Theta)
\end{equation}
and on the other hand
\begin{equation}
Q_{4,aaaa}(\vec{M}) = M_0^4 + M_1^4 - \frac{3}{4} (M_0^2 + M_1^2)^2
= \frac{1}{4} (M_0^4 - 6 M_0^2 M_1^2 + M_1^4) \;.
\end{equation}
The difference is that in Ref. \cite{Shao20} the magnetisation $\vec{M}$
is normalized for each configuration, while here we take first the
expectation value of $Q_{4,aaaa}(\vec{M})$ and normalize by using 
the expectation value of the magnetisation to the fourth 
power $\langle M^2 \rangle^2$.  The authors simulate at $h=1$.
There should be corrections due to the fact that they simulate at 
a finite value of $h$, similar to the finite $\mu$ effects here.
Furthermore there should be corrections proportional to $L^{-\omega}$.  
By chance, these different corrections seem to cancel in their 
estimate, which is consistent with ours.

\section{Acknowledgement}
This work was supported by the DFG under the grant No HA 3150/5-4.

\appendix
\section{The algorithm}
\label{algorithm}
The  Monte Carlo algorithm is similar to that of refs. 
\cite{myCubic,myCubic2}. It is a hybrid of local updates, the single cluster
algorithm \cite{Wolff} and the wall cluster \cite{KlausStefano} update.

The local update is needed to get an ergodic algorithm.
The local update allows to change the absolute value of the field. 
Since the reflection axis of the cluster update is restricted to four 
directions, as we discuss below, the local update is also needed to update 
the field variable to any angle.
In the local Metropolis algorithm, we generate a proposal by
\begin{equation}
\label{step}
 \phi_{x,i}' =  \phi_{x,i} + s (r_i -0.5)
\end{equation} 
for each component $i$ of the field at the site $x$. $r_i$ is a uniformly
distributed random number in $[0,1)$ and the step size $s$ is tuned such
that the acceptance rate is roughly $50 \%$. Note that for each component
a random number $r_i$ is taken.
The proposal is  accepted with the probability
\begin{equation}
\label{Pacc}
 P_{acc} = \mbox{min}\left[1,\exp(-H(\{\vec{\phi}\,\}') + H(\{\vec{\phi}\,\})\right]  \;.
\end{equation}

We have implemented over-relaxation updates
\begin{equation}
 \vec{\phi}_x^{\;\;'} =
 2 \frac{\vec{\Phi}_x \cdot \vec{\phi}_x}{\vec{\Phi}_x^2} \vec{\Phi}_x
- \vec{\phi}_x \;\;,
 \end{equation}
 where
\begin{equation}
\vec{\Phi}_x = \sum_{y.nn.x}  \vec{\phi}_y \;\;,
\end{equation}
where $\sum_{y.nn.x}$ is the sum over all nearest neighbors $y$ of the
site $x$.  In the case of the $O(N)$-invariant $\phi^4$ model
this update does not change the value of the Hamiltonian and
therefore no accept/reject step is needed.
Here, the value of the term $\mu \sum_{a} Q_{4, a a a a}(\vec{\phi}_x)$
changes under the update,
which has to be taken into account in an accept/reject step, eq.~(\ref{Pacc}).
This update has no parameter which can be tuned. The acceptance rate depends
on the parameters of the model. In particular, the larger $\mu$, the smaller
the acceptance rate. 

In our production runs, we used the over-relaxation update as a second hit
that follows the Metropolis update at a given site.

Let us turn to the cluster algorithms.
The field variable $\vec{\phi}_x$, where the site $x$ is in
the  cluster, is updated as 
\begin{equation}
\vec{\phi}\;'= \vec{\phi} -2 (\vec{r} \cdot \vec{\phi}) \vec{r} \;,
\end{equation}
where $\vec{r}$ is a two component unit vector. Note that $\vec{r}$
and $-\vec{r}$ are equivalent here.
In the cluster update, the local  potential~(\ref{localP})
should not change.
For finite $\mu$, this means that $\vec{r}$ is restricted
to the values $(1,0)$, $(2^{-1/2},2^{-1/2})$, $(0,1)$, and
$(-2^{-1/2},2^{-1/2})$.

$\vec{r} = (1,0)$ and $(0,1)$ flip the sign of the first or the second 
component of $\vec{\phi}$, respectively. This is very convenient in the 
simulation, since only one component of the field has to be touched
in the update. Combined with the local update, this is a valid algorithm, 
that also reduces critical slowing down considerably.
However, it turned out that using in addition 
$\vec{r} =(2^{-1/2},2^{-1/2})$ and $(-2^{-1/2},2^{-1/2})$, further 
reduces the autocorrelation times considerably.
Let us apply the corresponding updates on the field:
\begin{equation}
(\phi_0,\phi_1) - 2 [2^{-1/2} \phi_0 + 2^{-1/2} \phi_1](2^{-1/2},2^{-1/2})
= (\phi_0,\phi_1) -( \phi_0 +\phi_1,\phi_0 +\phi_1)= (-\phi_1,-\phi_0) \;,
\end{equation}
which means that the components are exchanged and the sign is reversed.
For the other value of $\vec{r}$ we get
\begin{equation}
(\phi_0,\phi_1) - 2 [-2^{-1/2} \phi_0 + 2^{-1/2} \phi_1](-2^{-1/2},2^{-1/2})
= (\phi_0,\phi_1) -( \phi_0 -\phi_1,-\phi_0 +\phi_1)= (\phi_1,\phi_0) \;,
\end{equation}
which means that the components are exchanged.
These two types of updates are performed by using the single cluster algorithm.
With probability $1/2$, we take either $\vec{r}=(2^{-1/2},2^{-1/2})$ or 
$(-2^{-1/2},2^{-1/2})$.

The update cycle is given by the following pseudo C-code:
\begin{verbatim}
for(k=0; k<3; k++)
{
metro_over();
for(jc=0; jc<L0; jc++) cluster_single();
metro_over();
cluster_wall(k,0); cluster_wall(k,1); 

measurements;
}
\end{verbatim}

Here \verb+k+ gives the direction of the wall, and the second argument
of \verb+cluster_wall+  the component that is updated. In 
\verb+metro_over()+ we perform a sweep with a two-hit update of the 
field variable. First a Metropolis update is performed, then at the 
same site the over-relaxation update follows.

We performed a few preliminary runs to determine the optimal step size $s$, 
eq.~(\ref{step}).  We take the lattice size $L=12$ and the two sets of
parameters $(\lambda,\mu)=(2.1,0.4)$ and $(1.5,1.3)$ simulating at
preliminary estimates of $\beta_c$. In both cases
we get the smallest autocorrelation times for $s \approx 1.8$, where the 
minimum is quite shallow. Hence an accurate fine tuning is not necessary.
For $s = 1.8$, we get $P_{acc} = 0.43987(2)$ and $0.46545(2)$ for 
$(\lambda,\mu)=(2.1,0.4)$  and $(1.5,1.3)$ respectively.
For the over-relaxation update, we get $P_{acc} = 0.956963(4)$ and
$0.85058(1)$, respectively.
In the following, in our production runs, we take $s=1.8$ in most cases. 

In table \ref{autocorr} we have summarized results for integrated
autocorrelation times $\tau_{int,A} $ for the three quantities
$E$, $\chi$ and $Q_4=\sum_{a} Q_{4,a a a a}$, eq.~(\ref{newterm}).  
Throughout $1.2 \times 10^6$ update cyles
after equilibration are performed. The simulations are performed for 
$(\lambda,\mu)=(1.8,1.0)$ at $\beta=0.498794$, close to our final estimate
of $\beta_c$.  The simulations were performed for three different 
compositions of the update cycle:
\begin{itemize}
\item   The update cycle given above (all)
\item   The update cycle give above, but the single cluster update taken out 
(wall) 
\item
the update cycle give above, but the wall cluster update
taken out (single)
\end{itemize}
Our numerical results are given in table \ref{autocorr}.

Furthermore, we have performed simulations with the local Metropolis plus 
over-relaxation update alone
for $L=12$, $16$, and $24$. As expected, the autocorrelation times 
rapidly increase, compatible with $\tau_{int} \propto L^{-z}$, 
where $z\approx 2$.  For example, for $L=12$, we get $\tau_{int,E}=44.4(1.1)$,
$\tau_{int,\chi} = 44.4(1.1)$ and $\tau_{int,Q_4} = 18.5(7)$.
Note that here, in contrast to the update cycle given above, 
there is only one \verb+metro_over()+ update sweep per measurement.

Adding the single cluster update, which either exchanges the components or 
exchanges the components and changes of the sign of both 
components, to the Metropolis update, we see that the autocorrelation time 
is already considerably reduced for $L=12$. Furthermore
the increase of $\tau_{int,A}$ is much slower than in the case of purely 
local updates. For the single cluster update combined with the Metropolis
update, the fastest increase of $\tau_{int,A}$ is observed for $Q_4$.
Fitting with the Ansatz $\tau_{int,Q_4}=a L^{-z}$  we get 
$z =0.58(3)$ taking into account the lattice sizes $L=32$, $48$ and $64$.

Combining the wall cluster update, which flips the sign of a single 
component, with the local Metropolis update, we get smaller autocorrelation
times than for the single cluster update. In particular, the integrated
autocorrelation time of $Q_4$ is considerably smaller.

But finally, combining the two different cluster updates, the integrated
autocorrelation times clearly further decrease. Taking into account the 
increased CPU time, still a clear advantage remains. In the case 
of $Q_4$ the integrated autocorrelation time remains roughly constant 
starting from $L=24$. In the case of $\chi$ the autocorrelation times 
are more than halved. The largest autocorrelation time is the one of the 
energy. Fitting these numbers with the Ansatz $\tau_{int,E}=a L^{-z}$ 
we get $z=0.370(6)$, taking into account all lattice sizes.

We also performed simulations, where we have replaced the wall cluster
updates with single cluster updates, which change the sign of a single
component of the field. Here we find a very similar result for the 
autocorrelation times. This shows that the combination of reflection 
axis is the key element, and not the particular type of the cluster 
update. 

Finally we note that while running the local update alone, adding the 
second hit with the overrelaxation update more than halves the 
autocorrelation times, there is virtually no effect, when combining 
with the cluster updates.

\begin{table}
\caption{\sl \label{autocorr}
Estimates of the integrated autocorrelation time for the energy $E$,
the magnetic susceptibility $\chi$ and $Q_4$ in units of measurements
for three different compositions of the update cycle discussed in the text.
$L$ is the linear lattice size. The simulations were performed at 
$(\lambda,\mu)=(1.8,1.0)$ at a preliminary estimate of $\beta_c$.
}
\begin{center}
\begin{tabular}{llll}
\hline
$L$ &   $E$    & $\chi$    &    $Q_4$   \\      
\hline
\mc{4}{c}{single} \\
\hline
12 &  5.73(5)  & 4.93(5)   &   3.50(5) \\
16 &  6.46(7)  & 5.58(6)   &   4.32(6)  \\
24 &  7.63(8)  & 6.85(7)   &   5.95(9)  \\
32 &  8.65(11) & 7.97(10)  &   7.24(12)  \\
48 & 10.34(14) & 10.16(15) &   9.89(20)  \\
64 & 11.73(18) & 12.21(19) &  11.99(23)  \\
\hline
\mc{4}{c}{wall} \\
\hline
12 & 5.15(5)  & 4.58(4) & 2.22(3)  \\
16 & 5.93(6)  & 5.25(5) & 2.69(3)  \\
24 & 7.13(7)  & 6.18(6) & 3.31(4) \\
32 & 8.18(9)  & 7.04(8) & 3.91(6) \\
48 & 9.95(13) & 8.34(11)& 4.48(7) \\
64 & 11.30(15)& 9.28(12)& 5.12(8) \\
\hline
\mc{4}{c}{all}  \\
\hline
12 &   3.37(3) & 2.72(2) & 1.45(2)   \\
16 &   3.74(3) & 2.91(3) & 1.55(2)  \\
24 &   4.41(4) & 3.25(3) & 1.62(2)  \\
32 &   4.89(4) & 3.43(3) & 1.66(2)  \\
48 &   5.60(5) & 3.73(3) & 1.62(2)  \\
64 &   6.25(6) & 4.00(4) & 1.64(2)  \\
\hline
\end{tabular}
\end{center}
\end{table}

\section{The improved one-component $\phi^4$ model}
\label{appendixB}
We discuss the model defined by the reduced Hamiltonian~(\ref{HamiltonianI}).
In the literature, the  
problem has been studied in \cite{Ball98,myPhi4} about 25 years ago.
In ref. \cite{Ball98} the authors quote $\tilde \lambda^*=1.0(1)$.  In
\cite{myPhi4} we quote
$\tilde \lambda^*=1.100(7)$, $1.102(8)$, and $1.095(12)$ from fits with 
$L_{min}=12$, $14$, and $16$, respectively. $L_{min}$ is the minimal linear
lattice size that is taken into account in the fit. The Ansatz used is
\begin{equation} 
\overline{U}_{4,I}(L,\tilde \lambda) = \overline{U}^*_{4,I} + c_1(\tilde \lambda) L^{-\omega} 
+ c_2 \left[c_1(\tilde \lambda) L^{-\omega}\right]^2  \;,
\end{equation}
where $\overline{U}_{4,I}$ is the Binder cumulant at a given value of 
$(Z_a/Z_p)_I$.

Here we improve the accuracy of the estimate by using moderate amount 
of CPU time. We make use of the accurate estimates of $R_i^*$ obtained 
recently \cite{myIso} from simulations of the improved Blume 
Capel model with nearest and next-to-next-to nearest neighbor 
couplings:
\begin{eqnarray}
(Z_a/Z_p)_I^* &=& 0.54253(1) \; , \\
U_{4,I}^* &=& 1.60359(4) \; , \\
U_{6,I}^* &=& 3.10535(10) \; ,\\
(\xi_{2nd}/L)_I^* &=& 0.64312(1) \; .
\end{eqnarray}
We performed new simulations at $\tilde \lambda=0.9$, $1.1$, and $1.3$. We simulated 
at the estimates of $\tilde \beta_c= 0.3845113(36)$, $0.3750966(4)$, and 
$0.3652233(33)$, respectively, obtained in ref. \cite{myPhi4}.
In the case of $\tilde \lambda=0.9$ and $1.3$ we simulated the linear lattice 
sizes $L=2$, $4$, $6$, $8$, $10$, $12$, $14$, and $16$. 

For $\tilde \lambda=1.1$ we simulated  the linear lattice sizes $L=2$, $3$, 
$4$, ..., $16$, $18$, ..., $24$, $28$, and $32$.  In the case of $L=32$
we performed $27906 \times 60000 \approx 1.7 \times 10^9$ measurements.
Down to $L=16$ the statistics is similar. For smaller lattice sizes it is
even larger.

The update cycle is given by
\begin{verbatim}
metro(); 
for(j=0;j<L0/4;j++) cluster_sing();
boundary();
measurement();
\end{verbatim}
For $L=32$, we used about the equivalent of 50 CPU days on a single core of a 
Intel(R) Xeon(R) CPU E3-1225 v3 running at 3.20 GHz.

\subsection{Analysis of the data}
Here we analyze $U_{4,I}$ at 
$(Z_a/Z_p)_I=0.54253$, which is denoted by $\overline{U}_{4,I}$.  
We reanalyzed the 
data of ref. \cite{myIso} and find, as for $U_{4,I}^*$, 
$\overline{U}_{4,I}^*=1.60359(4)$.  

\begin{figure}
\begin{center}
\includegraphics[width=14.5cm]{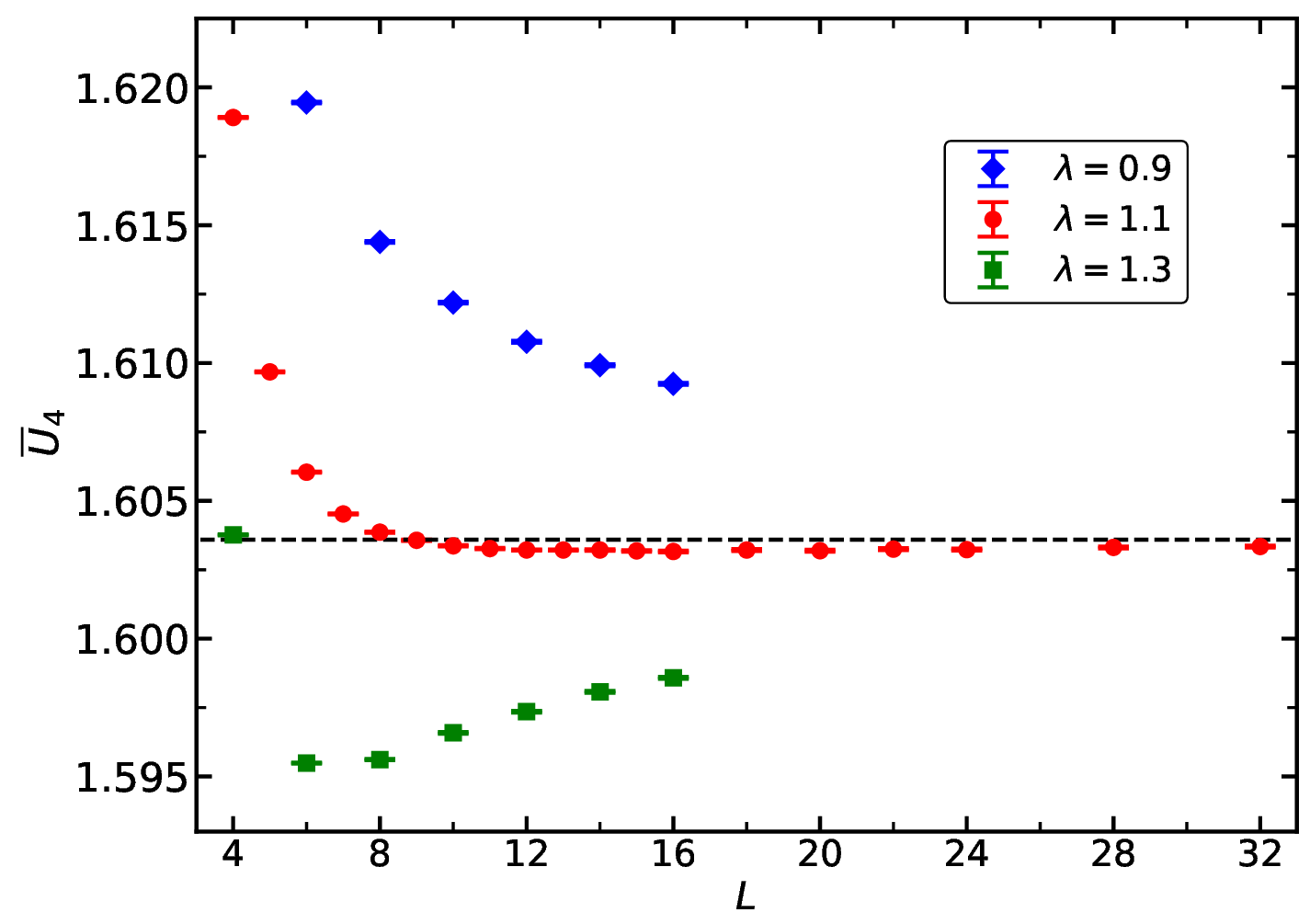}
\caption{\label{U4plot}
One component $\phi^4$ model.
We plot the Binder Cumulant $U_{4,I}$ at $(Z_a/Z_p)_I=0.54253$ as a 
function of the 
linear lattice size $L$ for $\tilde \lambda=0.9$, $1.1$ and $1.3$. For
comparison we give $\overline{U}_{4,I}^*=1.60359$ as dashed line.
}
\end{center}
\end{figure}

In Fig. \ref{U4plot}, we plot $\overline{U}_{4,I}$ versus the linear 
lattice size $L$. 
For very small values of $L$, subleading corrections to scaling numerically
dominate the deviation of $\overline{U}_{4,I}$ from $\overline{U}_{4,I}^*$.
Qualitatively it is clear from the plot that for $\tilde \lambda=0.9$ 
and $1.3$, 
for large $L$, $\overline{U}_{4,I}^*$ is approached from above and below, 
respectively. In the case of $\tilde \lambda=1.1$, $\overline{U}_{4,I}^*$ is 
undershoot starting from $L=9$. Still  $\overline{U}_{4,I}=1.603343(33)$
for $L=32$ is clearly below $\overline{U}_{4,I}^*$. 
Hence we expect that $\tilde \lambda^*$ is actually slightly smaller that 
$1.1$. In order to quantify these findings we performed fits of our data 
for $\tilde \lambda=1.1$ by using the Ans\"atze:
\begin{equation}
\label{Co1}
\overline{U}_{4,I}=\overline{U}_{4,I}^* + a L^{-\omega} \;,
\end{equation}
\begin{equation}
\label{Co2}
\overline{U}_{4,I}=\overline{U}_{4,I}^* + a L^{-\omega} + b L^{-2+\eta} \;
\end{equation}
and
\begin{equation}
\label{Co3}
\overline{U}_{4,I}=\overline{U}_{4,I}^* + a L^{-\omega} + b L^{-2+\eta} + c L^{-\omega'} \;,
\end{equation}
where we take $\omega=0.82968$, $2-\eta=1.9637022$, and $\omega'=3.8956$
\cite{SD16,CB_Ising_2024}. Here we abstain from fits with a correction 
exponent $\omega_{NR}=2.022665(28)$ \cite{SD16} due to the breaking of 
spatial rotational invariance by the lattice, assuming that it is effectively
taken into account by the correction due to the analytic background with
the exponent $2-\eta=1.9637022$.
We performed these fits, fixing $\overline{U}_{4,I}^*=1.60359$, and in order to 
estimate the error due to the uncertainty of our estimate of 
$\overline{U}_{4,I}^*$ to $\overline{U}_{4,I}^*=1.60363$.

\begin{figure}
\begin{center}
\includegraphics[width=14.5cm]{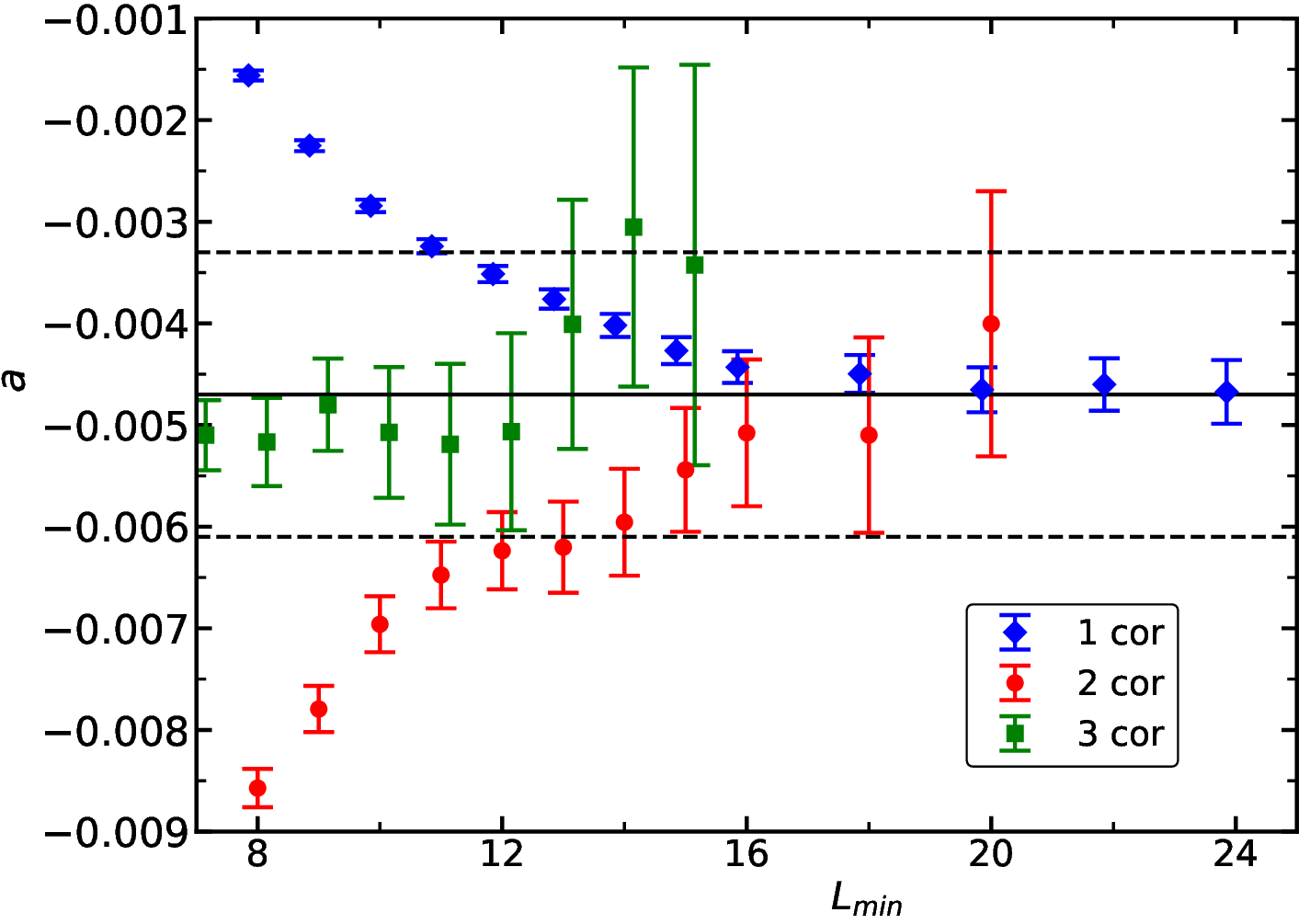}
\caption{\label{corAmp}
One component $\phi^4$ model. We plot estimates of the leading correction 
amplitude $c$ in $\overline{U}_{4,I}$ obtained by using the Ans\"atze 
(\ref{Co1},\ref{Co2},\ref{Co3}) using $\overline{U}_{4,I}^*=1.60359$. 
In the caption, these Ans\"atze are 
denoted by 1 cor, 2 cor, and 3 cor, giving the number of correction 
terms. The solid line gives our final estimate and the dashed lines 
indicate the error, assuming $\overline{U}_{4,I}^*=1.60359$. 
Note that the value of $L_{min}$ is slightly shifted
to avoid overlap of the symbols.
}
\end{center}
\end{figure}

In Fig. \ref{corAmp} we plot the estimates of the correction amplitude 
obtained by using the Ans\"atze (\ref{Co1},\ref{Co2},\ref{Co3}) and 
$\overline{U}_4^*=1.60359$. Our final estimate is taken such that it
is consistent at least with one of the estimates with an acceptable  
fit for each of the Ans\"atze.
For $\overline{U}_{4,I}^*=1.60359$ fixed, we get $a= -0.0047(14)$.
Taking into account the uncertainty of $\overline{U}_{4,I}^*$ we arrive at 
the final result
\begin{equation}
\label{am_result}
 a= -0.0047(21) \;.
\end{equation}

Next we analyzed the difference 
$\Delta=\overline{U}_{4,I}(\tilde \lambda=1.3)
-\overline{U}_{4,I}(\tilde \lambda=0.9)$. Taking 
the difference
$\overline{U}_{4,I}^*$ cancels and, as the numerics suggest, also subleading 
correction to a good extend.
Even the simplest fit $\Delta=d L^{-0.82968}$ gives acceptable 
$\chi^2/$DOF starting from $L_{min}=8$.  We arrive at the estimate 
$d=-0.106(2)$, taking also the results of more extended Ans\"atze 
into account.

Hence eq.~(\ref{am_result}) translates to the new estimate
\begin{equation}
\tilde \lambda^*= 1.1 
 - \left( -0.0047(21) \frac{0.4}{-0.106(2)} \right)=1.082(8) \;.
\end{equation}
Translating this to the DI case, we get, following Sec. \ref{DI_para},
$(\lambda_{DI}^*,\mu_{DI}^*)=(1.211(10),4 \lambda_{DI}^*/3)$.

\subsection{Consistency check of the program for the two-component model}
\label{consistency}
We simulated by using the program used for the two-component $\phi^4$
model at the parameters that correspond to  $\tilde \lambda =1.1$
and $\tilde \beta=0.3750966$  as derived in Sec.  \ref{DI_para}.
These are $\lambda=1.233725341045739$, $\mu=1.644967121394319$ and 
$\beta=0.458696467933495$. 
Since the local updates mix the two components and also the single cluster 
mixes the components, this is a nontrivial consistency check. 
Using the expressions for the dimensionless quantities given in section
\ref{DI_R}, we converted the numerical results  for $\tilde \lambda =1.1$
to the dimensionless quantities defined for the decoupled two-component system.
We performed simulations for $L=2$, $3$, $4$, $6$, $8$ and $10$. 
We get consistent results.
\section{The Ashkin-Teller model on the simple cubic lattice}
\label{appendixC}
The Ashkin-Teller model \cite{AshkinTeller} has been introduced in 1943.
Two copies of the Ising model are coupled via a four-point coupling.
The reduced Hamiltonian is given by
\begin{equation}
\label{Ashkin_Hamil}
 {\cal H} = -  \sum_{\left<xy\right>}  
  \left [K_2 ( 
          \sigma_x  \sigma_y + \tau_x  \tau_y)
         + K_4 \sigma_x  \sigma_y \tau_x  \tau_y \right] \;,
\end{equation}
where $\sigma_x, \tau_x \in \{-1,1\}$ and $\left<xy\right>$ denotes a pair of
nearest neighbor sites on the simple cubic lattice. For a meanfield solution 
see \cite{Kadanoff}. The model has a rich phase diagram. In three dimensions 
part of it is still disputed. A full discussion is far beyond the scope of 
this Appendix.
In two dimensions, there are rigors results. 
For recent work see for example Ref. \cite{AoDoGl24}.

Let us briefly discuss the part of the phase diagram of the three-dimensional 
model that is related to our study.
For $K_4=0$ there are two independent copies of the Ising model. 
The physics discussed here, is found in the neighborhood of $(K_2,K_4)=(K_{2,c},0)$, 
where $K_{2,c}$ is the critical coupling of the Ising model on the simple cubic 
lattice. For $K_4>0$, one gets first order phase transitions \cite{ArZh97}.
Note that in Ref. \cite{ArZh97} the simulation of Ashkin-Teller model by using 
the cluster algorithm  is nicely discussed. On the other hand for $K_4<0$ there
is a second order phase transition in the XY universality class, where the amount 
of the cubic perturbation depends on $K_4$. For recent work see Ref. \cite{ZhHuSuLv25}.

\def\refname{}

\end{document}